\documentclass[aps,prd,longbibliography ,twocolumn,nofootinbib,reprint]{revtex4-2}
\usepackage{graphicx}
\usepackage{amsmath}
\usepackage{amsfonts}
\usepackage{graphicx}
\usepackage{dsfont}
\usepackage[german, english]{babel}
\usepackage{amssymb}
\usepackage{ifthen}
\usepackage{ulem}
\usepackage{color}
\usepackage{float}
\usepackage{physics }
\usepackage[colorlinks=true,citecolor=blue,urlcolor=blue,linkcolor=blue]{hyperref}
\usepackage{bbold}
\usepackage{nccmath}
\usepackage{xcolor}

\newcommand{\da}{^\dagger}
\newcommand{\mycomment}[1]{}

\begin{document}
\title{A microscopic study of boundary superconducting states on a honeycomb lattice}
\begin{abstract}
    We address the problem of boundary s-wave superconductivity on rectangular honeycomb lattices: nanoflakes, armchair and zigzag nanotubes.
    We discuss how the presence of edges and corners in these systems can significantly alter the superconducting correlations at a macroscopic length scale, leading to either nontrivial enhancement or suppression of the superconducting gap value near the boundaries.
    This in turn results in different critical temperatures of the gap closure at boundaries compared to the bulk gap. The effects are macroscopic but strongly depend on the atomic-level structure of the boundaries. 
\end{abstract}
\author{Anton Talkachov}
\email{anttal@kth.se}
\author{Albert Samoilenka}
\author{Egor Babaev}
\affiliation{Department of Physics, KTH-Royal Institute of Technology, SE-10691, Stockholm, Sweden}
\date{\today}
\maketitle

\section{Introduction}
Recently the problem of superconductivity near the boundaries of a Bardeen–Cooper–Schrieffer (BCS) superconductor was revisited. The original calculations in BCS theory \cite{de1964boundary,de1966superconductivity,CdGM_Coherence,CdGM_french,abrikosov1965concerning}
came to the conclusion that the superconducting gap approaches the surface of a BCS superconductor with zero normal derivative.
It was shown in \cite{samoilenka2020boundary,samoilenka2020pair,benfenati2021boundary,barkman2022elevated,samoilenka2021microscopic,hainzl2022boundary} that instead surfaces, corners and edges of a BCS superconductor have in general higher critical temperature than the bulk. The effect is closely connected with the oscillation of density of states near boundaries, allowing to construct a highly inhomogeneous solutions of the gap equation that have higher critical temperature than nearly-uniform solutions. Although the theoretical results also indicated that the effect is strongly dependent on surface quality and hence can be modified by oxidation or different chemical composition of the surface \cite{samoilenka2020boundary,barkman2022elevated}, nonetheless 
there are experimental reports on boundary superconductivity \cite{fink1969surface,lortz2006origin, janod1993split,khlyustikov2011critical,khlyustikov2016surface,kozhevnikov2007observation,khlyustikov2021surface,mangel2020stiffnessometer,tsindlekht2004tunneling,belogolovskii2010zirconium,khasanov2005anomalous}.
The previous theoretical studies were primarily focused on the cases of simplest square or rectangular lattices or continuum theories. That rises the question of the interplay between these effects and the existence of nontrivial localized single-electron states
on different lattices. One of the very simplest example one can consider is the case of a honeycomb lattice that has nontrivial boundary states \cite{nakada1996edge, fujita1996peculiar, wakabayashi1999electronic, Wakabayashi_2010, PhysRevB.71.193406, PhysRevB.73.045124,shtanko2018robustness}.

To study the interplay between these effects we consider the problem of boundary and bulk critical temperatures on a honeycomb lattice. While the realization of various unconventional superconducting pairing symmetries were proposed for such lattices (for a review see \cite{pangburn2022superconductivity}) our goal is to compare the effects of different symmetries of the lattice on the boundary effects in \cite{samoilenka2020boundary,samoilenka2020pair,benfenati2021boundary,barkman2022elevated,samoilenka2021microscopic,hainzl2022boundary} and to that end, we consider the case of the simplest s-wave pairing interaction within mean-field approximation.
Although the considerations would apply also to other systems with similar lattice effects, for brevity, below we refer to the honeycomb system as graphene.

\section{Infinite structure} \label{section: infinite}

Let us first look at the infinite honeycomb structure made from identical atoms. We divide these atoms into two groups $(A, B)$ to form two sublattices. Effective Hubbard Hamiltonian for the system reads
\begin{equation}
\label{eq:effective_Hamiltonian}
\begin{split}
    H_\text{eff} = &-t\sum_{\langle \textbf{i},\textbf{j} \rangle} \sum_{\sigma = \uparrow, \downarrow} {\left( a_{\textbf{i},\sigma}\da b_{\textbf{j},\sigma} + b_{\textbf{j},\sigma}\da a_{\textbf{i},\sigma} \right)} \\
    &- \mu \sum_{\textbf{i}} \sum_{\sigma = \uparrow, \downarrow} {\left( a_{\textbf{i},\sigma}\da a_{\textbf{i},\sigma} + b_{\textbf{i},\sigma}\da b_{\textbf{i},\sigma} \right)} \\
    &-V \sum_{\textbf{i}} {\left( a_{\textbf{i},\uparrow}\da a_{\textbf{i},\uparrow} a_{\textbf{i},\downarrow}\da a_{\textbf{i},\downarrow} +    b_{\textbf{i},\uparrow}\da b_{\textbf{i},\uparrow} b_{\textbf{i},\downarrow}\da b_{\textbf{i},\downarrow}  \right)}.
\end{split}
\end{equation}
Here $a_{\textbf{i},\sigma}\da (a_{\textbf{i},\sigma})$ is creation (annihilation) operator for electron with spin $\sigma$ on site $A$ in cell which position is described with vector $\textbf{i}=(n, m)$, where $n$ ($m$) specifies horizontal (vertical) position. The same applies to operators $b_{\textbf{i},\sigma}\da$ and $b_{\textbf{i},\sigma}$ which correspond to sites $B$.
In Eq. (\ref{eq:effective_Hamiltonian}) first term describes kinetic energy (hopping between nearest-neighbour sites $\langle \textbf{i},\textbf{j} \rangle$ without spin flip), parameterized by the hopping integral $t$ ($t>0$).
The second term associated with chemical potential $\mu$ controls filling. The last term describes attraction energy between electrons in the same site using potential $V$ ($V>0$). 
All parameters $t$, $\mu$, $V$ are assumed to be constant in space. 
The main focus of this work will be on the physics of boundaries and boundary superconducting state that was recently discussed 
on square lattices and on continuum \cite{samoilenka2020boundary,samoilenka2021microscopic,barkman2022elevated,barkman2019surface,benfenati2021boundary,samoilenka2020pair}.
In order to compare with the previously considered cases, here we focus on s-wave pairing.
Further, all energies, $\mu$, $V$, and temperature $T$ are measured in the units of $t$ for simplicity.

\begin{figure}
    \centering
    \includegraphics[width=0.99\linewidth]{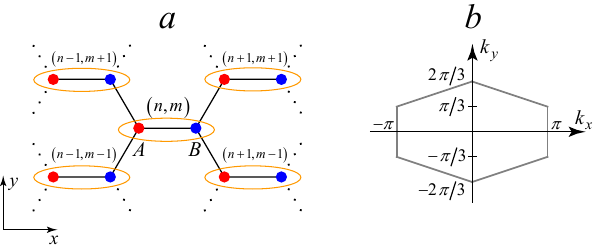}
    \caption{(a) Honeycomb lattice in real space, where the red (blue) circles mean an $A$ ($B$)-sublattice site. (b) 1st Brillouin zone for the Bloch's theorem expansion Eq. (\ref{eq:Bloch_theorem}).}
    \label{fig: honeycomb infinite structure}
\end{figure}

We apply the Hartree–Fock–Bogoliubov mean-field approximation. The transformed one-particle mean-field Hamiltonian reads
\begin{equation}
\label{eq:mean-field_Hamiltonian}
\begin{split}
    H_\text{MF} = &-\sum_{\langle \textbf{i},\textbf{j} \rangle} \sum_{\sigma} \left( a_{\textbf{i},\sigma}\da b_{\textbf{j},\sigma} + b_{\textbf{j},\sigma}\da a_{\textbf{i},\sigma} \right) \\
    & - \mu \sum_{\textbf{i}} \sum_{\sigma} {\left( a_{\textbf{i},\sigma}\da a_{\textbf{i},\sigma} + b_{\textbf{i},\sigma}\da b_{\textbf{i},\sigma} \right)} \\
    &+ \sum_{\textbf{i}} \Bigl( \Delta_{\textbf{i},A} a_{\textbf{i},\uparrow}\da a_{\textbf{i},\downarrow}\da +
    \Delta_{\textbf{i},B} b_{\textbf{i},\uparrow}\da b_{\textbf{i},\downarrow}\da \\
    & + \Delta^{*}_{\textbf{i},A} a_{\textbf{i},\downarrow} a_{\textbf{i},\uparrow} + \Delta^{*}_{\textbf{i},B}  b_{\textbf{i},\downarrow} b_{\textbf{i},\uparrow} \Bigr) + \text{const},
\end{split}
\end{equation}
where introduced superconducting mean-field order parameter $\Delta_{\textbf{i},\text{type}}$ (here type means $A$ or $B$ sublattice)
\begin{equation}
\label{eq:delta_definition}
    \Delta_{\textbf{i},A} = -V \langle a_{\textbf{i},\downarrow} a_{\textbf{i},\uparrow} \rangle, \qquad
    \Delta_{\textbf{i},B} = -V \langle b_{\textbf{i},\downarrow} b_{\textbf{i},\uparrow} \rangle.
\end{equation}
This parameter is constant in space in the case of an infinite system.

Mean-field Hamiltonian is quadratic and it can be diagonalized with the following Bogoliubov transformation for a unit cell consisting of two atoms:
\begin{equation}
\label{eq:Bogoliubov_transformation}
    \mqty(a_{\textbf{i},\sigma} \\ b_{\textbf{i},\sigma}) = \sum_\nu^{'} \mqty(u_{\textbf{i}}^\nu \\ y_{\textbf{i}}^\nu) \gamma_{\nu, \sigma} - 
    \sigma \sum_\nu^{'} \mqty(v_{\textbf{i}}^{\nu *} \\ z_{\textbf{i}}^{\nu *}) \gamma_{\nu, -\sigma}\da.
\end{equation}
Here, operator $\gamma_{\nu, \sigma}\da$ $(\gamma_{\nu, \sigma})$ creates (annihilates) a quasiparticle in the state $\nu$ with the spin $\sigma$ ($\sigma = \uparrow = 1$, $\sigma = \downarrow = -1$), prime sign means summation over states with positive excitation energy. These operators satisfy the standard anticommutation relations $\{ \gamma_{\nu, \sigma},\gamma\da_{\nu^{'}, \sigma^{'}} \} = \delta_{\nu, \nu^{'}}\delta_{\sigma, \sigma^{'}}$,
$\{ \gamma_{\nu, \sigma},\gamma_{\nu^{'}, \sigma^{'}} \} = \{ \gamma\da_{\nu, \sigma},\gamma\da_{\nu^{'}, \sigma^{'}} \} = 0$.
Diagonalized Hamiltonian reads
\begin{equation}
    H_\text{MF} = E_g + \sum_\nu^{'} \sum_\sigma E^\nu \gamma_{\nu, \sigma}\da \gamma_{\nu, \sigma},
\end{equation}
where $E_g$ is ground state energy. $E^\nu$ are excitation energies (we are looking for $E^\nu > 0$) which can be obtained from the following system of Bogoliubov–de Gennes equations with self-consistent conditions:
\begin{equation}
\label{eq:BdG_eq}
    \sum_\textbf{j}
    \begin{pmatrix}
        H_0 (\textbf{i},\textbf{j}) && \Delta (\textbf{i},\textbf{j}) \\
        \Delta\da (\textbf{i},\textbf{j}) && -H_0^* (\textbf{i},\textbf{j})
    \end{pmatrix}
    \mqty(u^{\nu}_\textbf{j} \\ y^{\nu}_\textbf{j} \\ v^{\nu}_\textbf{j} \\ z^{\nu}_\textbf{j})
    =
    E^\nu \mqty(u^{\nu}_\textbf{i} \\ y^{\nu}_\textbf{i} \\ v^{\nu}_\textbf{i} \\ z^{\nu}_\textbf{i}),
\end{equation}
\begin{equation}\label{eq:delta_in_vectors}
\begin{gathered}
\Delta_{\textbf{i},A} = V \sum_\nu^{'} u_{\textbf{i}}^\nu v_{\textbf{i}}^{\nu *} \tanh \frac{E^\nu}{2 T}, \\
\Delta_{\textbf{i},B} = V \sum_\nu^{'} y_{\textbf{i}}^\nu z_{\textbf{i}}^{\nu *} \tanh \frac{E^\nu}{2 T}.
\end{gathered}
\end{equation}
where $H_0 (\textbf{i},\textbf{j})$ and $\Delta (\textbf{i},\textbf{j})$ are $2 \times 2$ matrices. Explicit forms of the matrices and derivation of the self-consistent conditions are given in Appendix \ref{app:ap1}.

The eigenvalue problem (Eq. (\ref{eq:BdG_eq})) can be significantly simplified in the limit of the infinite size of the system.
Due to transnational and rotational symmetries $\Delta_{\textbf{i},A} = \Delta_{\textbf{i},B} = \Delta$ in this case.
We use the translational symmetry in both $x$ and $y$ directions because the order parameter is constant in the above-mentioned limit.
Applying Bloch's theorem one can expand eigenvectors in plane waves for $\textbf{i} = (n,m)$:
\begin{equation}
\label{eq:Bloch_theorem}
    \mqty(u^{\nu}_\textbf{i} \\ y^{\nu}_\textbf{i} \\ v^{\nu}_\textbf{i} \\ z^{\nu}_\textbf{i})
    = \frac{1}{\sqrt{N_x N_y / 2}} \sum_{k_x, k_y} e^{i (k_x n + k_y m)}
    \mqty(\mathcal{U}_\textbf{k} \\ \mathcal{Y}_\textbf{k} \\ \mathcal{V}_\textbf{k} \\ \mathcal{Z}_\textbf{k}),
\end{equation}
where $N_x$ ($N_y$) is number of atoms in $x$ ($y$) direction, $k_x$ and $k_y$ are wavenumbers which located in the first Brillouin zone (1st BZ).
This Brillouin zone (Fig. \ref{fig: honeycomb infinite structure}b) is halved in $k_x$  direction and compressed $2 / \sqrt{3}$ times in $k_y$ direction in comparison to the conventional choice of Brillouin zone for honeycomb lattice (which has a shape of regular hexagon with radius $4 \pi / 3$ for the choice of unit length between nearest sites).
Its area $S_\text{1st BZ} = 2 \pi^2$. 
Here $k_y$ has $N_y$ different values, $k_x$ has $N_x /2$ values because in $x$ direction unit cell that we chose consists of 2 atoms.

Substituting Eq. (\ref{eq:Bloch_theorem}) to Eq. (\ref{eq:BdG_eq}) and solving matrix equation one can obtain eigenvalues $E_s$:
\begin{equation}
\label{eq:energies with delta}
    E_s = \pm \sqrt{\epsilon^{2}_s + \Delta\Delta^*},
\end{equation}
\begin{equation}
\label{eq:energies}
\begin{gathered}
    \epsilon_s = -\mu + s \cdot \epsilon_0 (k_x, k_y),
    \\
    \epsilon_0 (k_x, k_y) = \sqrt{3 + 4 \cos{k_x} \cos{k_y} + 2 \cos{2 k_y}},
\end{gathered}
\end{equation}
where we introduced auxiliary functions $\epsilon_s$ and parameter $s = \pm 1$.
One can obtain well-known self-consistent condition with integration over the first Brillouin zone by switching from summation to integration (for the detailed derivation see Appendix \ref{app:ap2}):
\begin{equation}
    \frac{1}{V} = \frac{1}{4 S_{\text{1st BZ}}} \sum_{s = \pm 1} \iint_{\text{1st BZ}} dk_x dk_y \frac{\tanh{\frac{E_s}{2 T}}}{E_s}.
\end{equation}
This equation contains an implicit temperature dependence of the energy gap.
It can be further simplified ($\Delta \rightarrow 0$) to find critical temperature $T_{c1}$:
\begin{equation} \label{eq:infinite system integral}
    \medmath{\frac{1}{V} = \frac{1}{4 S_{\text{1st BZ}}} \iint_{\text{1st BZ}} dk_x dk_y} \left( \frac{\tanh{\frac{\epsilon_+}{2 T_{c1}}}}{\epsilon_+} + \frac{\tanh{\frac{\epsilon_-}{2 T_{c1}}}}{\epsilon_-} \right).
\end{equation}

\begin{figure}
    \centering
    \includegraphics[width=0.99\linewidth]{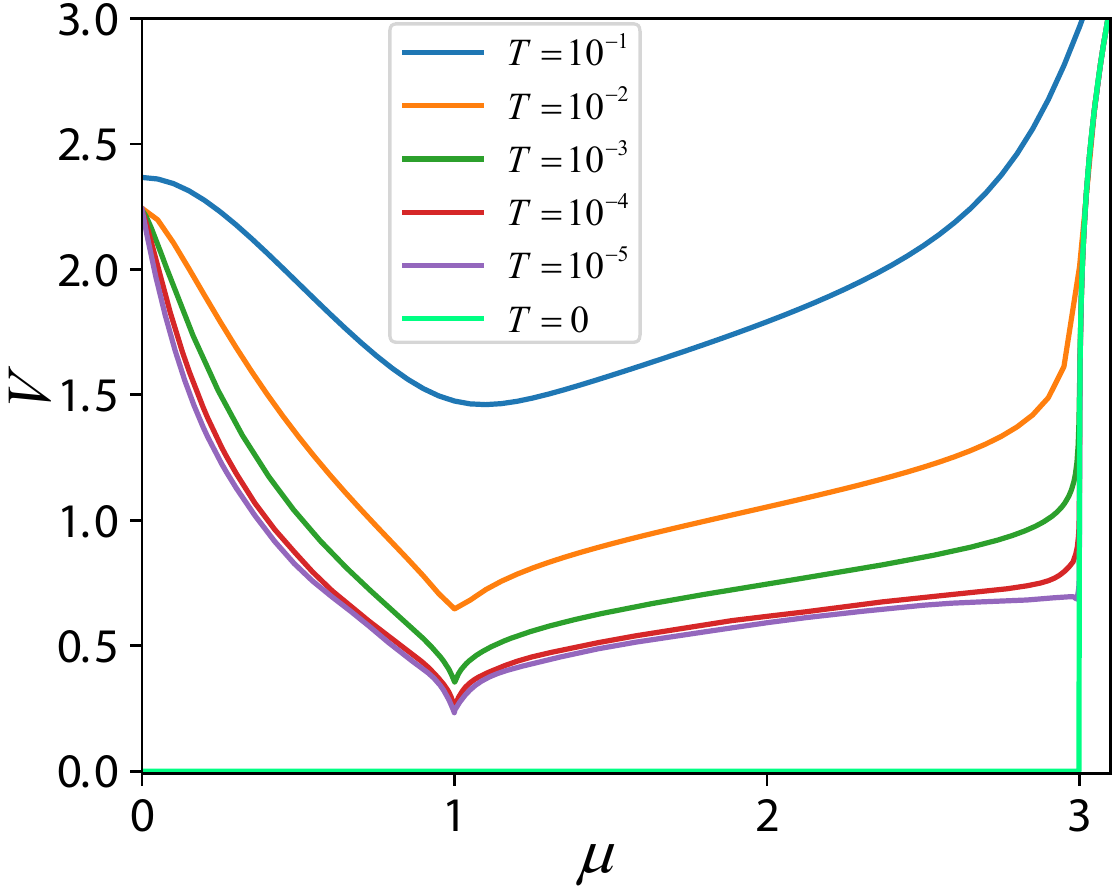}
    \caption{Infinite honeycomb lattice superconductivity phase diagram in chemical potential--attraction onsite potential coordinates for different critical temperatures. For a given temperature above the transition line, gap is nonzero and vice versa.}
    \label{fig: phase diagram}
\end{figure}

This equation allows us to calculate the superconductivity phase diagram (with $V$ and $\mu$ axes): find the transition between superconducting ($\Delta \neq 0$) and normal ($\Delta = 0$) states.
The phase diagram is shown in Fig. \ref{fig: phase diagram}, where superconductivity exists above-chosen transition line.
Cooling the system leads to decreasing critical pairing in the region $|\mu| \in [0; 3)$, but from Fig. \ref{fig: phase diagram} one can see that it is definitely nonlinear dependence.
One can ask two basic questions:
\begin{itemize}
    \item Is there a lower boundary for the curve (how does it look at $T=0$)?
    \item How does this curve approach zero temperature configuration?
\end{itemize}

Integral in Eq. (\ref{eq:infinite system integral}) was calculated numerically to obtain results in Fig. \ref{fig: phase diagram}.
Decreasing temperature leads to increasing numerical errors due to narrowing the region of energies ($|\epsilon_s| \lesssim T$) with the biggest contribution to the integral, so the questions can't be answered using a numerical approach. One can analytically show (see Appendix \ref{app:ap3}) that dominant contribution for the integral in Eq. (\ref{eq:infinite system integral}) close to zero temperature will be

\begin{equation}
    \frac{1}{V} \propto - \ln{T}
\end{equation}
for $\mu \in (0; 3)$. This tendency is also verified numerically with the result that it holds for $|\mu| \in (0; 1) \cup (1; 3)$, $T < 0.01$ and for higher temperatures when $|\mu|$ is far from exceptional points 0, 1, 3.

It also answers the first question by showing that we have $V=0$ boundary at $T=0$ in the region of chemical potential where Fermi surface has nonzero length ($|\mu| \in (0; 3)$).

\section{Finite systems}

\subsection{Linearized gap equation approach}

\begin{figure}
    \centering
    \includegraphics[width=0.99\linewidth]{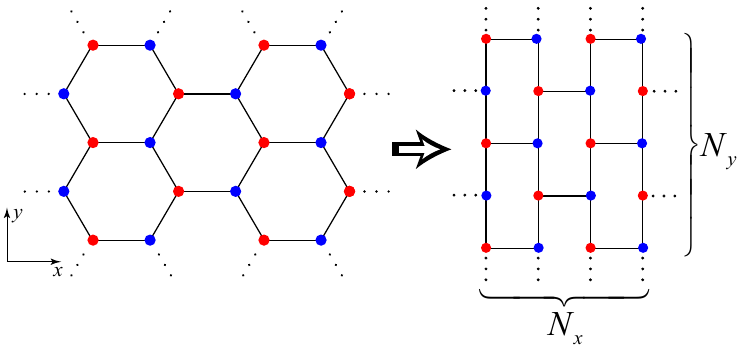}
    \caption{Honeycomb lattice transformation to a rectangular shape.}
    \label{fig: lattice transformation}
\end{figure}

The method we used to find the critical temperature in the previous section works only for an infinite structure where $\Delta$ is constant. 
Consider now the problem of calculation of $T_c$ for finite structures without an assumption of constant  $\Delta$. When the superconducting transition is second order at mean-field level (all $\Delta_\textbf{i} \rightarrow 0$ when $T \rightarrow T_c$) one can write Bogoliubov–de Gennes equations (\ref{eq:BdG_eq}) up to the leading order in $\Delta$:

\begin{equation}
\label{eq:Tc_discrete}
    \frac{1}{V} \Delta_{\textbf{i},\text{type}} = \sum_{\textbf{i}',\text{type}'} K_{\textbf{i},\text{type},\textbf{i}',\text{type}'} \Delta_{\textbf{i}',\text{type}'},
\end{equation}
\begin{equation}
\begin{gathered}
\label{eq:K_matrix}
    K_{\textbf{i},\text{type},\textbf{i}',\text{type}'} = \sum_{s,s'} \sum_{\textbf{k},\textbf{k}'} \frac{1 - f(\epsilon_s (\textbf{k}))- f(\epsilon_{s'} (\textbf{k}'))}{\epsilon_s (\textbf{k}) + \epsilon_{s'} (\textbf{k}')} \\
    \cdot w_{s,\textbf{k}}^* (\textbf{i},\text{type}) w_{s',\textbf{k}'}^* (\textbf{i},\text{type}) w_{s,\textbf{k}} (\textbf{i}',\text{type}') w_{s',\textbf{k}'} (\textbf{i}',\text{type}'),
\end{gathered}
\end{equation}
where $f(E)$ is the Fermi distribution function ($f(E) = (1+e^{E/{T}})^{-1}$), $w_n$ are the one-electron wave functions in the normal state (when $\Delta = 0$) corresponding to eigenenergies $\epsilon_n$.
They can be found in many papers \cite{Wakabayashi_2010,Saroka2017Optics,talkachov2022wave}.
Here summation over $\textbf{i}',\text{type}'$ means summation over all system sites, summation over $\textbf{k}$ means summation over all allowed  $k_x$ and $k_y$, $\epsilon_s$ are eigenenergies in a normal state defined in (\ref{eq:energies}).
If the system (Fig. \ref{fig: lattice transformation}\textit{a}) has $N_x$ atoms in the horizontal direction (along the armchair edge) and $N_y$ atoms in the vertical direction (along zigzag edge) matrix $K_{\textbf{i},\text{type},\textbf{i}',\text{type}'}$ has $N_x N_y \cross N_x N_y$ dimensions. 
Equation (\ref{eq:Tc_discrete}) is an eigenvalue problem: the largest eigenvalue of $K$ matrix gives $V^{-1}$ and the corresponding eigenvector is the energy gap distribution close to superconducting transition.

Let us apply the approach first to graphene nanotubes and then to finite rectangular graphene systems.

\subsection{Graphene nanotubes} \label{section: Graphene nanotubes}

Let us consider nanotubes with open armchair (periodic in $x$ direction, Fig. \ref{fig: lattice transformation}) and zigzag (periodic in $y$ direction) edges.
Further, we call them armchair and zigzag nanotubes respectively.
Free electron wave functions for the first case are extended states which are described by sine functions \cite{Wakabayashi_2010, wakabayashi2012nanoscale, Onipko2018Revisit, zheng2007analytical, talkachov2022wave}.
However, a zigzag nanotube has both extended and localized wave functions \cite{wakabayashi2012nanoscale,talkachov2022wave}.
Localized ones are called edge states and are described by exponents which describe the localization of the states near boundaries.
The number of edge states equals $N_y/3$ in the limit of wide ($N_x \gg 1$) zigzag nanotube \cite{nakada1996edge,talkachov2022wave}. Hence, the relative amount of edge states is $(3 N_x)^{-1}$ of the total number of states.

\begin{figure*}
    \centering
    \includegraphics[width=0.99\linewidth]{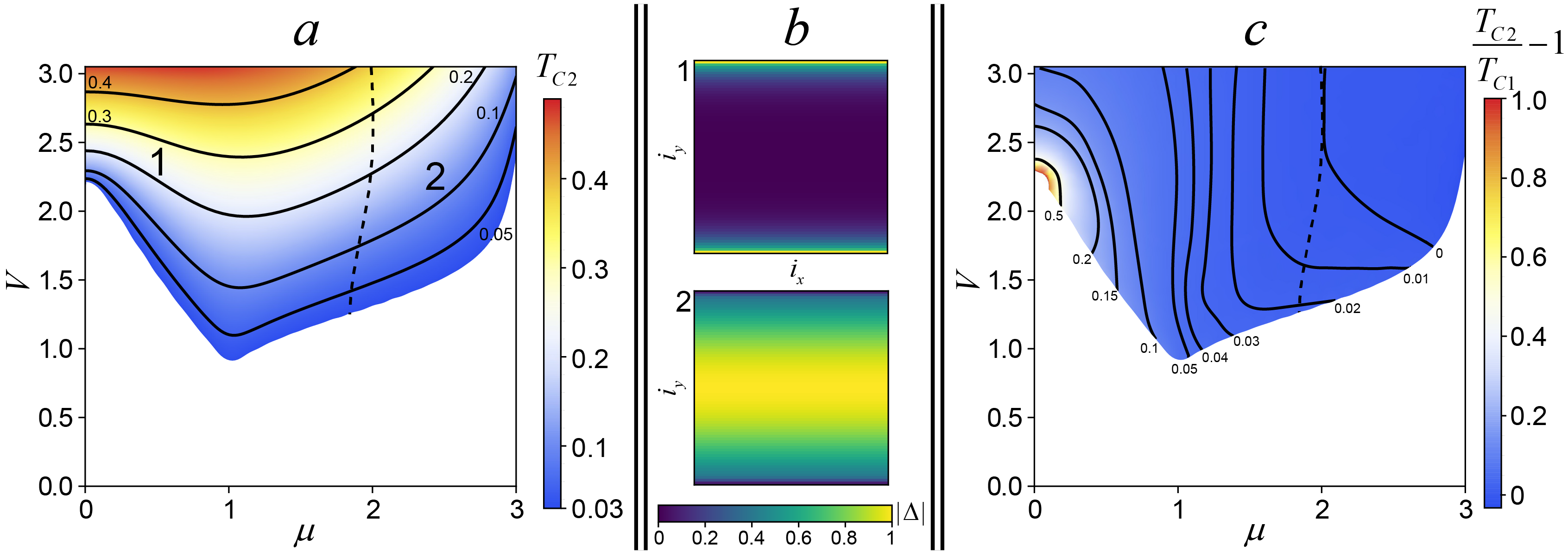}
    \caption{(\textit{a}) Phase diagram for an armchair nanotube (the system is periodic in $x$ direction and free in $y$ direction).
    Solid lines correspond to constant critical temperature curves with $T_c$ written close to the line.
    Big numbers 1 and 2 numerate the regions with different order parameter distributions illustrated in the part (\textit{b}).
    The dashed line is the 'transition line' between the two regions ($\Delta$ in the bulk and on the boundary are equal).
    (\textit{c}) Relative change in the $T_c$ for an armchair nanotube in comparison to the infinite graphene sheet.
    Solid lines are constant-level curves.
    The dashed line is the same as in the part (\textit{a}).}
    \label{fig: x periodic}
\end{figure*}

We employ linearized gap equation approach (Eqs. (\ref{eq:Tc_discrete}), (\ref{eq:K_matrix})) to examine superconducting phase transition in the two types of nanotubes.
Wave functions and eigenenergies are used from the Ref. \cite{talkachov2022wave}.
System size we used varied from $40 \cross 40$ ($8.4 \cross 4.8$ nm) to $70 \cross 70$ ($14.8 \cross 8.5$ nm).
Calculation of $K$ matrix (Eq. (\ref{eq:K_matrix})) is a computationally expensive problem because it scales as $\mathcal{O} [(N_x N_y)^4]$.
The other system sizes ($N_x \neq N_y$) were also studied with identical to the case $N_x = N_y$ result.
System size effect for the above mentioned systems range, $\forall \mu$, $T > 0.1$ is less than $0.1 \%$ (in $V$).
The effect is more significant for smaller systems.
For temperatures less than $0.05$ system size effect is noticeable even for $40 \cross 40$ systems.
This manifests itself in the form of oscillations superimposed on the overall trend of $V(\mu)$ function.
It can be seen on the bottom of Figs. \ref{fig: x periodic}\textit{a}, \ref{fig: y periodic}\textit{a}.
The main reason is the following: a 'weight function' (fraction in Eq. (\ref{eq:K_matrix})) is localized in the region $|\epsilon_s (\textbf{k})|, |\epsilon_{s'} (\textbf{k}')| \lesssim T$, density of states discretizes for a lattice, therefore, a smooth shift in chemical potential leads to a step-like change in the amount of non-zero values of $K$ matrix and consequently to significant change in the eigenvalue which is proportional to $V^{-1}$.
The amount of non-vanishing values of $K$ matrix is big for high $T$ and discrete change in the amount doesn't have a significant effect.
The lowest investigated temperatures are set to be 0.03 and 0.04 for armchair and zigzag nanotubes respectively.
The lowest temperatures are chosen as temperatures when the above-mentioned oscillations are visibly detected.
They are different for armchair and zigzag nanotubes due to the different density of states.
Calculations of density of states for infinite nanoribbons (it is the same as infinite radius nanotubes) show peaky structure \cite{Wakabayashi_2010}, therefore increasing system size won't solve the problem for the low temperatures.
Here we didn't discuss the influence of wave functions in Eq. (\ref{eq:K_matrix}) because they are temperature-independent.

As a check of our results we employed a self-consistent approach using spectral decomposition of Bogoliubov–de Gennes equations (\ref{eq:BdG_eq}) with Chebyshev polynomials \cite{weisse2006kernel, covaci2010efficient, nagai2012efficient} up to order 2000.
It allows us to calculate order parameter distribution for a given set of $\mu$, $V$, $T$.
We used it in the following way:
Using half-division method we are looking for the $V$ value which gives the largest $\Delta \in [10^{-5}; 10^{-4}]$ in the sample after 1000 iterations of self-consistent equation (\ref{eq:delta_in_vectors}) for given $\mu$ and $T$.
The method allows us only to estimate transition $V$ for given $\mu$ and $T$ because we don't achieve full convergence.
It always gives us a lower boundary for $V$, which is a few percent lower than $V$ values found from the linearized gap equation for $T>0.1$.
Temperature growth leads to a decrease in the difference.
However, the spectral Chebyshev polynomial decomposition approach also fails for the low temperatures due to the influence of Gibbs oscillations \cite{gibbs1899fourier}.

\subsubsection{Nanotubes with armchair boundary}

Figure \ref{fig: x periodic}\textit{a} shows a phase diagram of the superconducting phase transition for an armchair nanotube.
Here the critical temperature is called $T_{c2}$ because in general, it differs from the $T_{c1}$ for an infinite sample.
We separated the diagram into two regions (1 and 2) where one can note different distributions of the order parameter.
In the first region, $\Delta$ on the boundaries (top and bottom of the sample, because system is periodic in \textit{x} direction and open in \textit{y} direction) is higher than $\Delta$ in the center of the sample and the second region with the opposite criterion.
It doesn't mean that on the dashed line distribution of the order parameter is uniform (at the line sites with maximal $\Delta$ are located close to the boundary).

Figure \ref{fig: x periodic}\textit{b} shows the typical order parameter distributions (normalized to unity) in the regions.
Here we used square lattice representation by lattice transformation (Fig. \ref{fig: lattice transformation}).
One can see that in the first region the largest gap lies on the boundary, however, in the second region, it lies in the center.
One can describe boundary gap enhancement in region 1 as an exponentially decaying function $\Delta(i_y) \propto (e^{-y / \xi} + e^{(y - L_y) / \xi})$, where $\xi (\mu, T)$ is a coherence length, $L_y$ is the nanotube length.
This function works badly on the boundaries due to the presence of short-range oscillations (Wilbraham-Gibbs phenomenon \cite{wilbraham1848cambridge, gibbs1899fourier} which is also called Friedel oscillations), but can describe tails that overlap in the bulk.
Fitting the function to obtained gap distributions (like in Fig. \ref{fig: x periodic}\textit{b}) one can come to the following conclusion: $\xi (\mu, T)$ is an increasing function of $\mu$ and a decreasing function of $T$ in the region 1.
In region 2 boundaries lead to suppression of the gap which can be described by a similar function.
Here the coherence length $\xi (\mu, T)$ is a decreasing function of both parameters.

The relative change in the critical temperature in comparison to the infinite system (Eq. (\ref{eq:infinite system integral})) is shown in Fig. \ref{fig: x periodic}\textit{c}.
Here we restricted the maximal value to 1 ($100 \%$).
Almost the whole investigated region has $T_{c2} > T_{c1}$ which means that superconductivity in the armchair nanotube is enhanced in comparison to the infinite graphene sheet.
Combining the result with gap distributions (Fig. \ref{fig: x periodic}\textit{b}) we can say that superconductivity survives on the boundaries.
However, one can come to the opposite result (boundaries suppress superconductivity) for big values of $\mu$ (almost filled band).
One can see that increase in chemical potential leads to a monotonic decrease of relative change in $T_c$ and finally leads to negative values.

\subsubsection{Nanotubes with zigzag boundary}

\begin{figure*}
    \centering
    \includegraphics[width=0.8\linewidth]{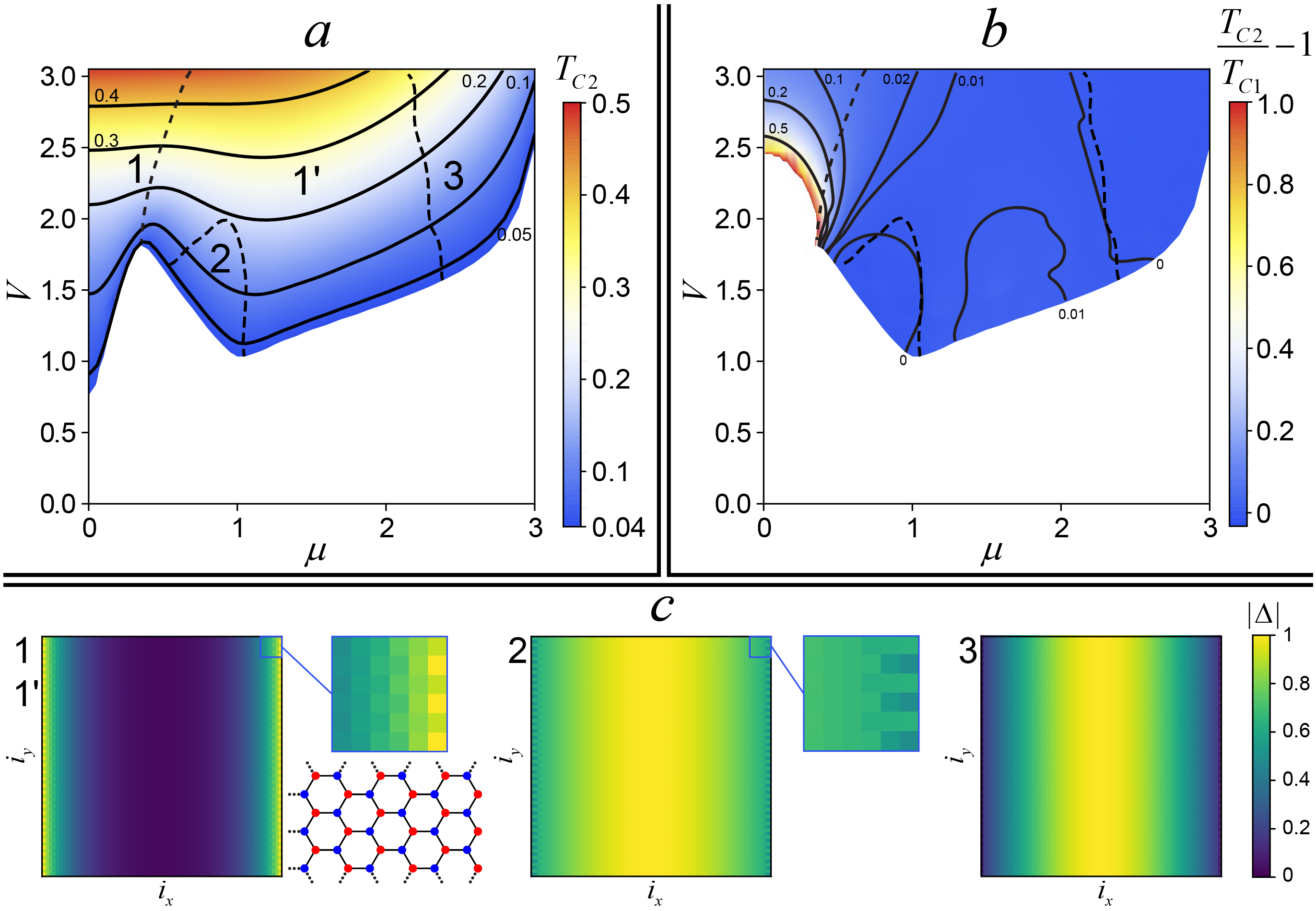}
    \caption{(\textit{a}) Phase diagram for a zigzag nanotube (the system is periodic in $y$ direction and free in $x$ direction).
    Solid lines correspond to constant critical temperature curves with $T_c$ written close to the line.
    Big numbers enumerate the regions with different order parameter distributions illustrated in the part (\textit{c}).
    Dashed lines are the 'transition lines' between the regions ($\Delta$ in the corresponding locations are equal).
    (\textit{b}) Relative change in the $T_c$ for a zigzag nanotube in comparison to the infinite graphene sheet.
    Solid lines are constant-level curves.
    Dashed lines are the same as in the part (\textit{a}).}
    \label{fig: y periodic}
\end{figure*}

Now we switch to the discussion of a zigzag nanotube which is a sample that periodic in \textit{x} direction and open in \textit{y} direction (Fig. \ref{fig: y periodic}).
Here one can note a significant change in the behaviour in the region $|\mu| \lesssim 0.5$ (Fig. \ref{fig: y periodic}\textit{a}), where $V$ is an increasing function of $\mu$ and lies lower than for the infinite graphene sheet (Fig. \ref{fig: phase diagram}) and the armchair nanotube (Fig. \ref{fig: x periodic}\textit{a}).
In this case, gap is not uniform in $y$ direction, because of the zigzag boundary, where only half of the 'boundary' atoms have two neighbours (Fig. \ref{fig: y periodic}\textit{c} where zoom is shown for $6 \times 6$ boundary region).
Again we divide the whole phase diagram into a few regions.
Here in regions 1 and 1' the average gap on the boundary is bigger than in the center (Fig. \ref{fig: y periodic}\textit{c}).
In region 1, the order parameter in the center is less than 0.001 (after gap normalization).
Regions 2 and 3 have the biggest gap in the center (Fig. \ref{fig: y periodic}\textit{c}).
We decided to call them differently because of their wide separation in the parameter space.
If one defines a boundary as atoms that have an absent neighbour (like red atoms on the right side in Fig. \ref{fig: y periodic}\textit{c}), regions 2 and 3 will slightly change the size and shape without qualitative differences.
Another reason to divide regions 2 and 3 is the quantitative gap suppression on the boundaries (Fig. \ref{fig: y periodic}\textit{c}): in region 2 it drops only to half of the $\Delta$ in the bulk, however, in region 3 the suppression is one order higher.

Analyzing typical gap distributions for zigzag nanotube (Fig. \ref{fig: y periodic}\textit{c}), one can note that enhancement (in regions 1, 1') or suppression (in regions 2, 3) origins from the boundary atoms which have two neighbours.
The exceptionality of the atoms is also underlined in the wave functions \cite{talkachov2022wave} where there are two zero energy states with non-zero wave function only at the sites.
There are also approximately $N_y / 3 - 2$ edge states which have almost zero energy.
They are the main reason for the significant difference between the zigzag nanotube phase diagram and previously considered systems.

Relative change in the critical temperature in comparison to an infinite graphene sheet is shown in Fig. \ref{fig: y periodic}\textit{b}.
Here we also restricted the maximal value to 1.
In this case in the region $|\mu| \lesssim 0.4$ there is a great increase in $T_c$ which can achieve order of hundreds that correspond to the edge localized nonzero gap states.
In region 1', the typical increase in critical temperature has an order of $1\%$.
Note, the regions with a decrease of $T_c$ (in comparison to the infinite sample) which almost fully overlap with regions 2 and 3.
In contrast to the armchair nanotube where the dashed line (which corresponds to equal $\Delta$ on boundaries and in the center) correlates with the line $T_{c1} = T_{c2}$ only in a small region of $V$ (Fig. \ref{fig: x periodic}\textit{c}).
Note that the cross-section for constant $V$ has non-monotonic behaviour in the relative change in $T_c$ (Fig. \ref{fig: y periodic}\textit{b}).

\subsubsection{LDOS argument for nanotubes}

Boundaries modify edge LDOS and it causes a change in $T_c$ in the region.
In the subsection, we investigate the interplay between the LDOS and superconductivity.

Thermalized LDOS at energy $E$ for the non-interacting model can be calculates as \cite{zhu2016bogoliubov}
\begin{equation}
    \text{LDOS}_i (E) = - \sum_{s, \textbf{k}} |w_{s, \textbf{k}}(i)|^2 f' \left(E - \epsilon_s(\textbf{k}) \right),
\end{equation}
where energies $\epsilon_s(\textbf{k})$ are defined in Eq. (\ref{eq:energies}). In BCS theory \cite{de1966superconductivity}  the bulk critical temperature for an infinite sample proportional to $\exp (- (V \cdot \text{DOS})^{-1})$. Here we consider local critical
temperature and  substitution DOS $\rightarrow$ LDOS for non-interacting system.
We note that boundary superconductivity is a complex phenomenon with many factors and direct substitution of LDOS is not necessarily sufficient for the assessment of the situation because it can oscillate at length scales much smaller than superconducting coherence lengths leading to nontrivial solutions 
\cite{samoilenka2020boundary,barkman2022elevated}.

\begin{figure}
    \centering
    \includegraphics[width=0.99\linewidth]{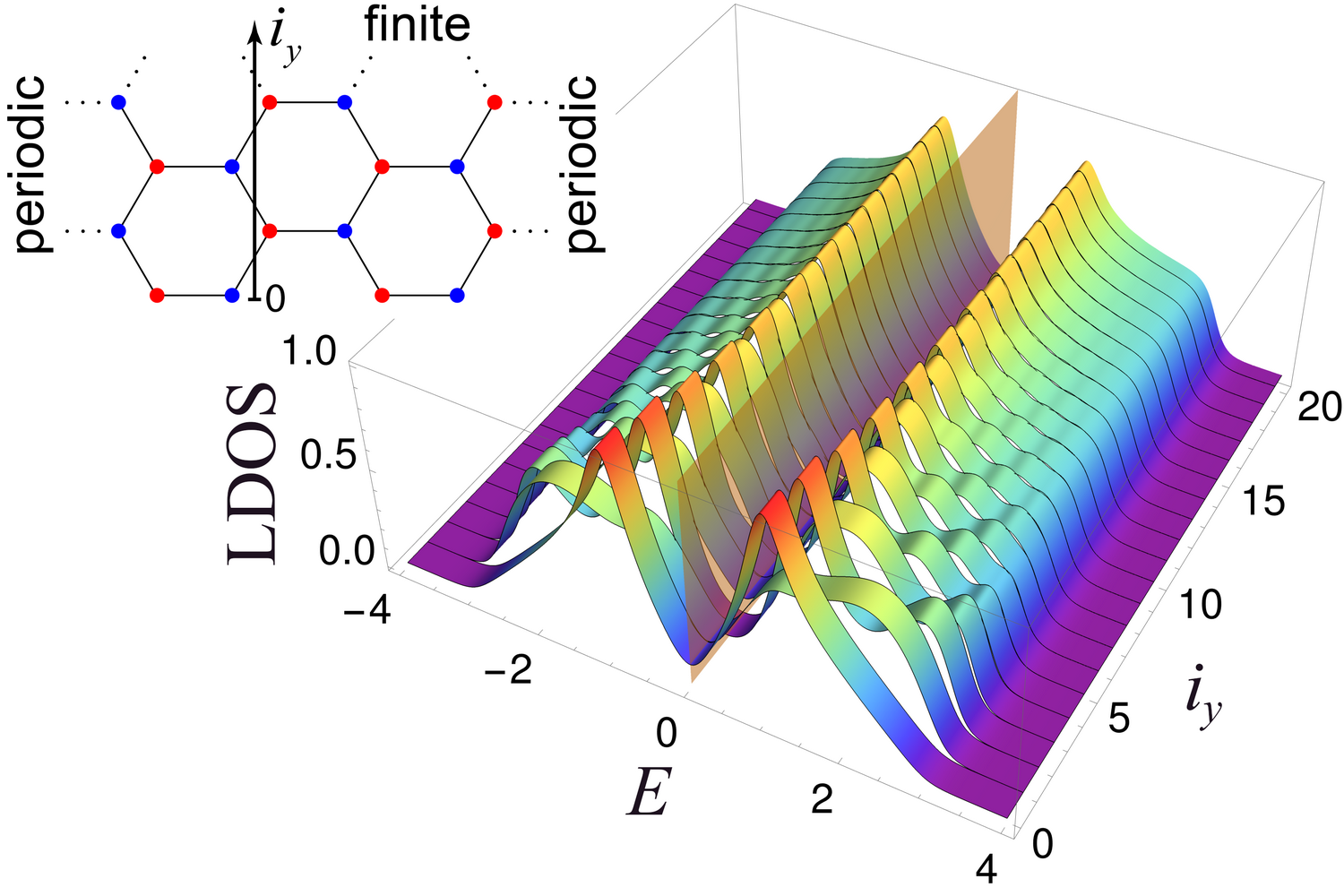}
    \caption{LDOS for an armchair nanotube without interaction for $\mu = 0$, $T_c = 0.1$. The orange plane corresponds to the Fermi level.}
    \label{fig: xperiodic LDOS}
\end{figure}

\begin{figure}
    \centering
    \includegraphics[width=0.99\linewidth]{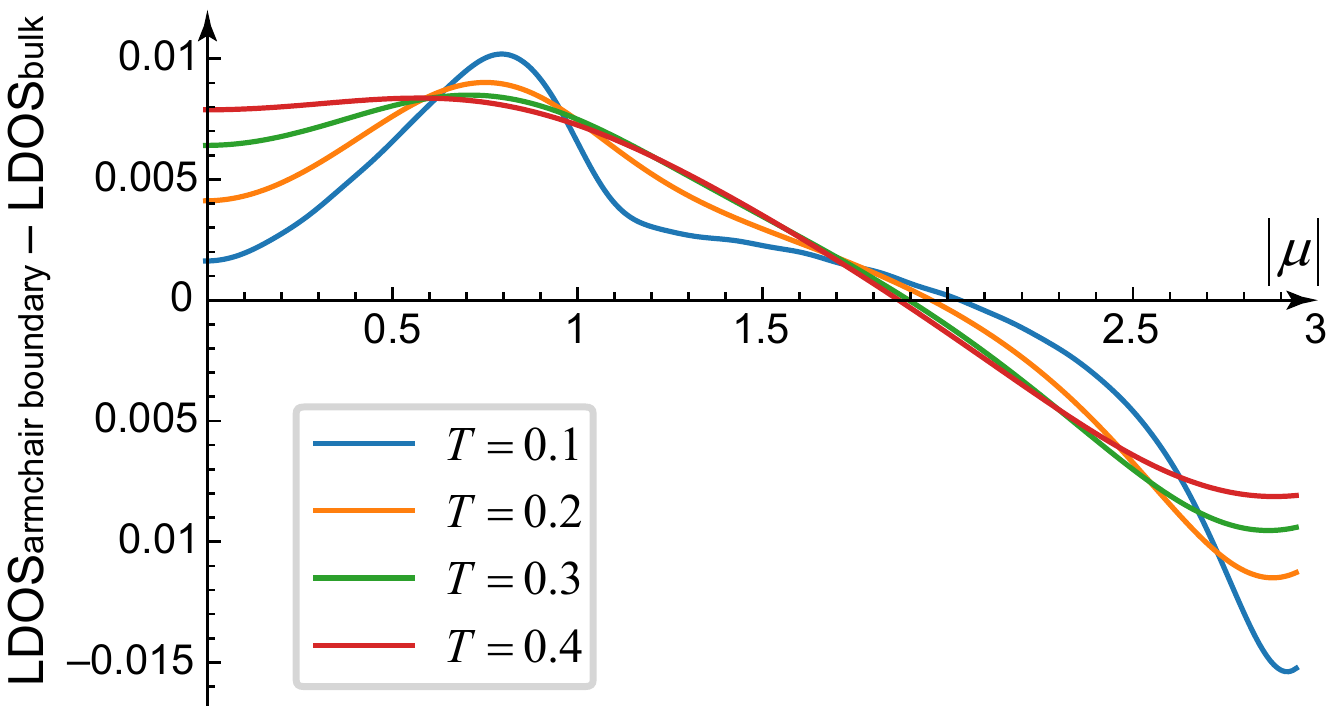}
    \caption{Difference between averaged LDOS at fifteen boundary sites of an armchair nanotube and LDOS in the bulk of the system as a function of chemical potential}
    \label{fig: xperiodic LDOS - bulk}
\end{figure}

There is only one unique direction parallel to the nanotube axis for an armchair nanotube (Fig. \ref{fig: xperiodic LDOS}).
LDOS in the direction at half-filling ($\mu = 0$) and $T_c = 0.1$ is shown in Fig. \ref{fig: xperiodic LDOS}.
It is normalized by the maximal LDOS value in the sample.
Here one can see significant deviations from bulk DOS (which can be seen for the large $i_y$ values in Fig. \ref{fig: xperiodic LDOS}) in the ten sites adjacent to the boundary.
Taking into account chemical potential just shifts the picture on $E$ axis by $\mu$.
Note that LDOS on the boundary sites varies when moving from the boundary. Let us consider average LDOS on $i_y \in [0; 14]$ (Fig. \ref{fig: xperiodic LDOS}).
Figure \ref{fig: xperiodic LDOS - bulk} shows the difference between the averaged boundary LDOS and bulk LDOS as a function of chemical potential for different temperatures.
Here one can see that the point where the difference is zero (LDOSes are equal) moves to smaller $\mu$ values when $T$ increases.
The line with equal critical temperatures for an armchair nanoribbon and the infinite sample is noted by '0' in Fig. \ref{fig: x periodic}\textit{c}.
Increasing the temperature (moving upwards along the '0' line) leads to a decrease in chemical potential.
The LDOS model (Fig. \ref{fig: xperiodic LDOS - bulk}) captures qualitative behaviour, however quantities of $\mu$ differ by 8--15 $\%$ from the values in Fig. \ref{fig: x periodic}\textit{c}.

\begin{figure*}
    \centering
    \includegraphics[width=0.99\linewidth]{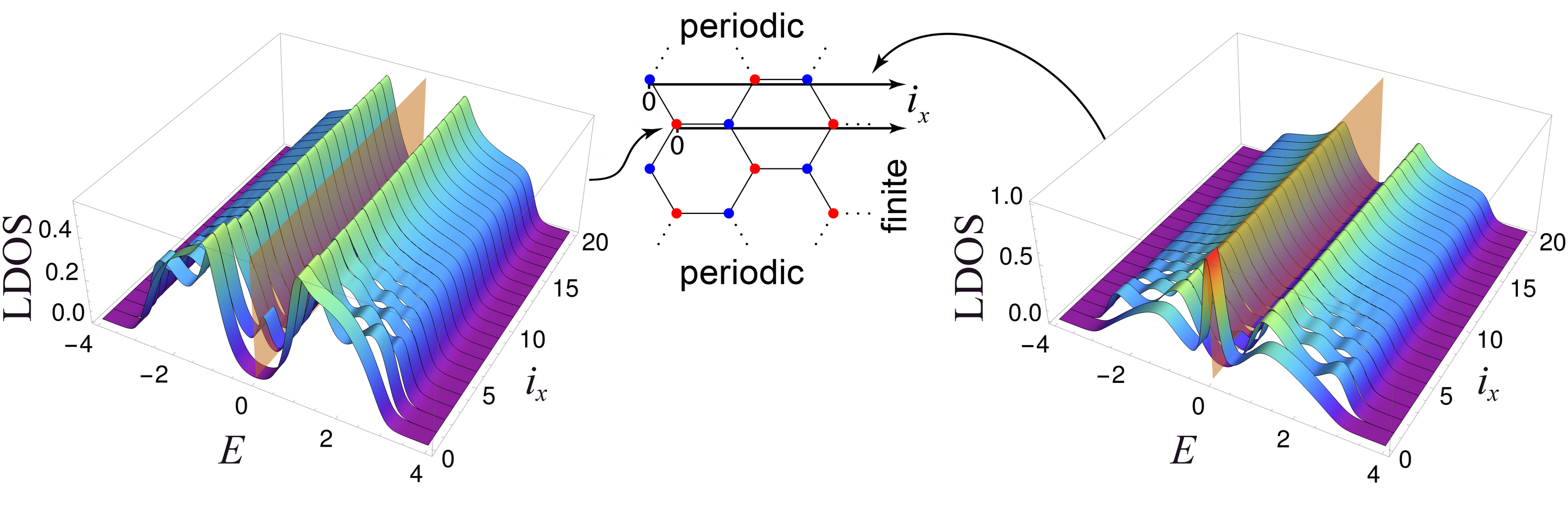}
    \caption{LDOS for a zigzag nanotube without interaction for $\mu = 0$, $T_c = 0.1$. The orange plane corresponds to the Fermi level.}
    \label{fig: yperiodic LDOS}
\end{figure*}

\begin{figure}
    \centering
    \includegraphics[width=0.99\linewidth]{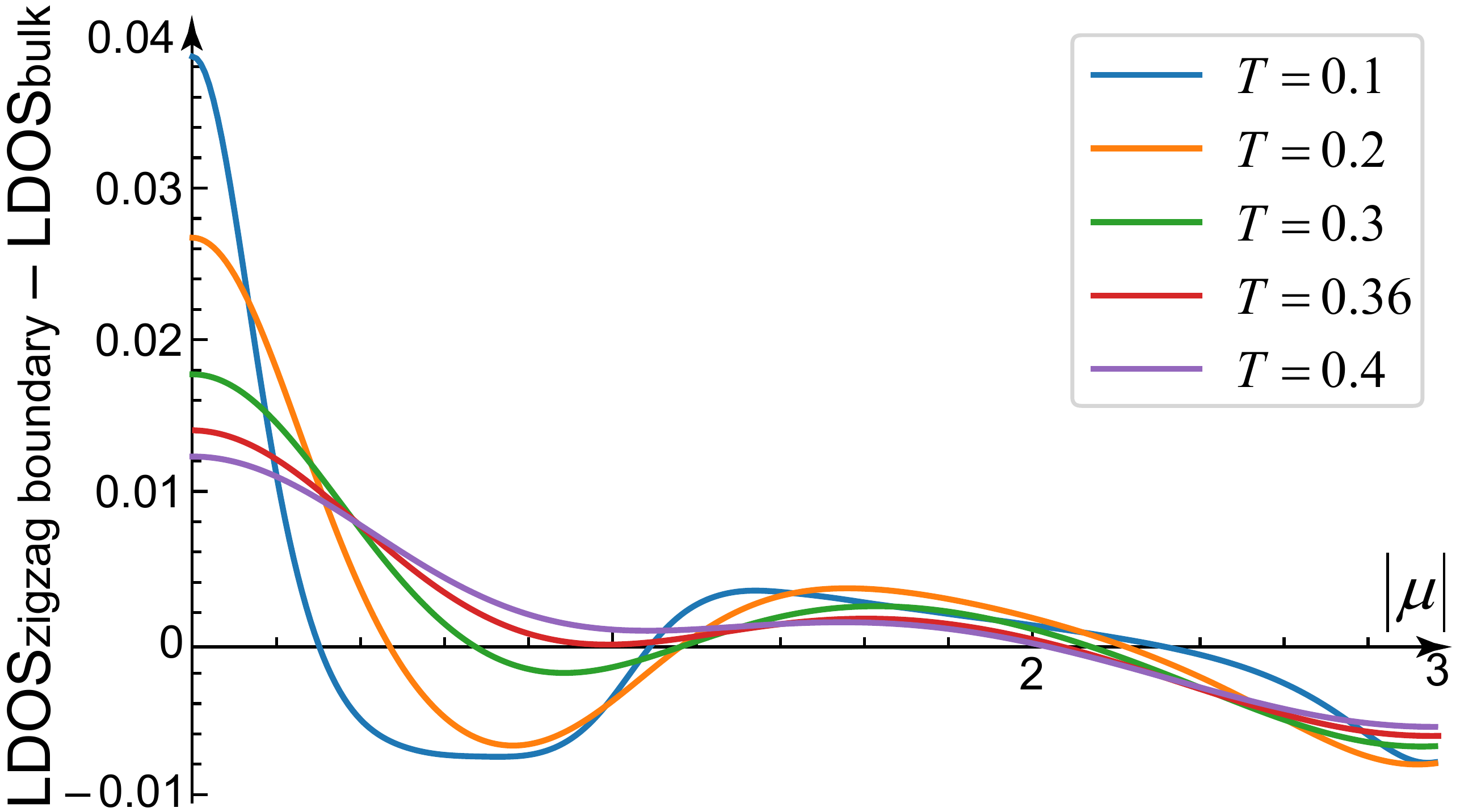}
    \caption{Difference between averaged LDOS at fifteen boundary sites (in two directions from Fig. \ref{fig: yperiodic LDOS}) of a zigzag nanotube and LDOS in the bulk of the system as a function of chemical potential}
    \label{fig: yperiodic LDOS - bulk}
\end{figure}

Now we apply the method for a zigzag nanotube.
There are two unique directions parallel to the nanotube axis for a zigzag nanotube (Fig. \ref{fig: yperiodic LDOS}).
In one of the directions, the atom at site 0 is not a true boundary atom because it has all three bonds.
The atom has LDOS similar to the bulk (Fig. \ref{fig: yperiodic LDOS} which is normalized by the maximal LDOS value in the sample).
In the other direction boundary atom LDOS is completely different from the bulk LDOS (Fig. \ref{fig: yperiodic LDOS}).
The reason is the existence of edge states with close to zero energy which are localized close to the boundaries \cite{wakabayashi1999electronic,wakabayashi2012nanoscale,Saroka2017Optics,talkachov2022wave}.
In the case of a zigzag nanotube, approximately five sites adjacent to the boundary have different LDOS from bulk LDOS which is twice smaller region in comparison to an armchair nanotube.
Figure \ref{fig: yperiodic LDOS - bulk} shows the difference between boundary LDOS (averaged over fifteen adjacent to the boundary sites in each of the directions) and bulk LDOS as a function of $\mu$.
We remind, that if the value is greater than zero, it means that we have a zigzag edge state, otherwise we have a bulk state.
In Fig. \ref{fig: yperiodic LDOS - bulk} one can see qualitative similarity with Fig. \ref{fig: y periodic}\textit{c}: region $\mu \in (0.3; 1.2)$ with negative LDOS difference values for $T < 0.36$ correspond to the region 2 in Fig. \ref{fig: y periodic}\textit{a} (where maximal $T_c = 0.19$).
The second similarity is the quantitative concurrence of the boundary between regions 1' and 3 in Fig. \ref{fig: y periodic}\textit{a} and the points for $\mu \in (2; 2.4)$ (Fig. \ref{fig: yperiodic LDOS - bulk}) where LDOS difference equals to zero.
The relative difference is less than 2 $\%$.

\subsection{Graphene rectangular finite samples}

\begin{figure}
    \centering
    \includegraphics[width=0.99\linewidth]{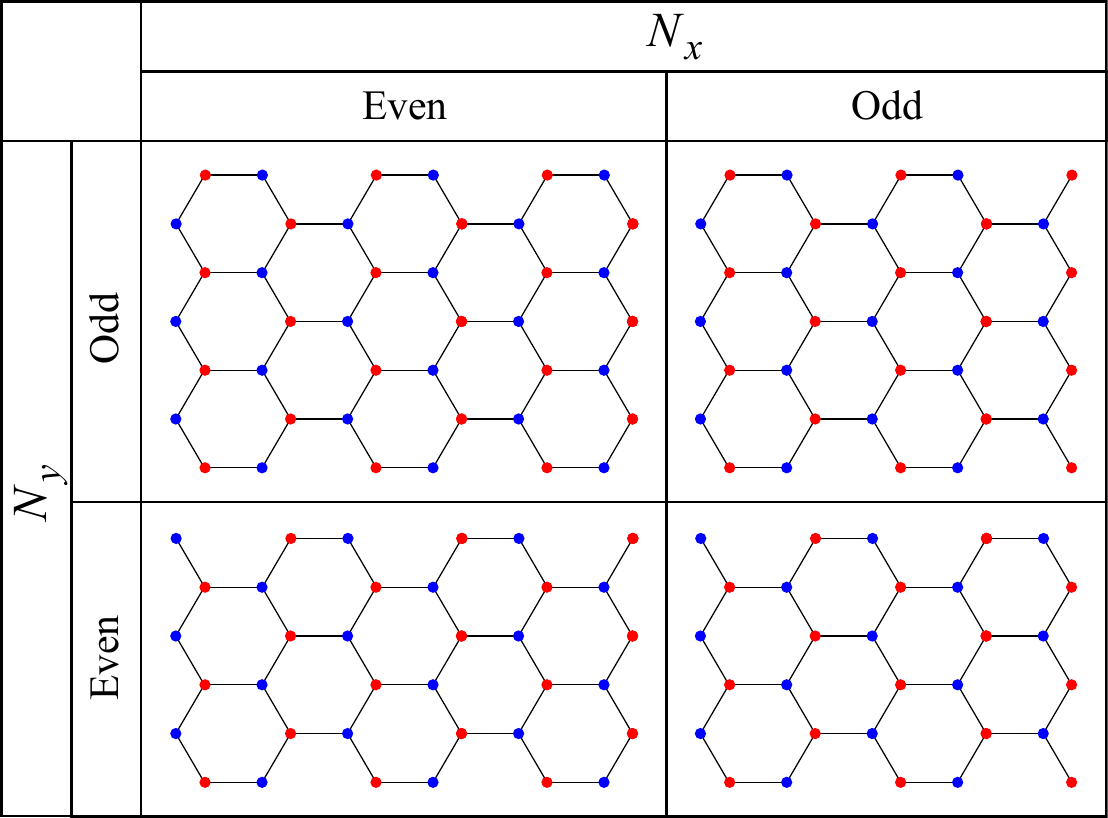}
    \caption{Possible rectangular geometries of the finite size honeycomb lattice.}
    \label{fig: finite structures}
\end{figure}

There are four possible finite rectangular graphene geometries (Fig. \ref{fig: finite structures}).
One of them (even $N_x$ and odd $N_y$) has a 'closed structure' which means that each atom has at least two neighbours.
Three other geometries have two atoms which have only one neighbour.
We carried out a similar to the previous section investigation of the four structures.
The result is that the three geometries have qualitatively and quantitatively similar phase diagrams which differ from the results for the 'closed structure'.
Therefore, first, we consider the case of even $N_x$ and odd $N_y$ geometry and then switch to the three other cases which will be discussed in the example of even both $N_x$ and $N_y$ geometry.

\subsubsection{The 'closed structure' case}

\begin{figure*}
    \centering
    \includegraphics[width=0.8\linewidth]{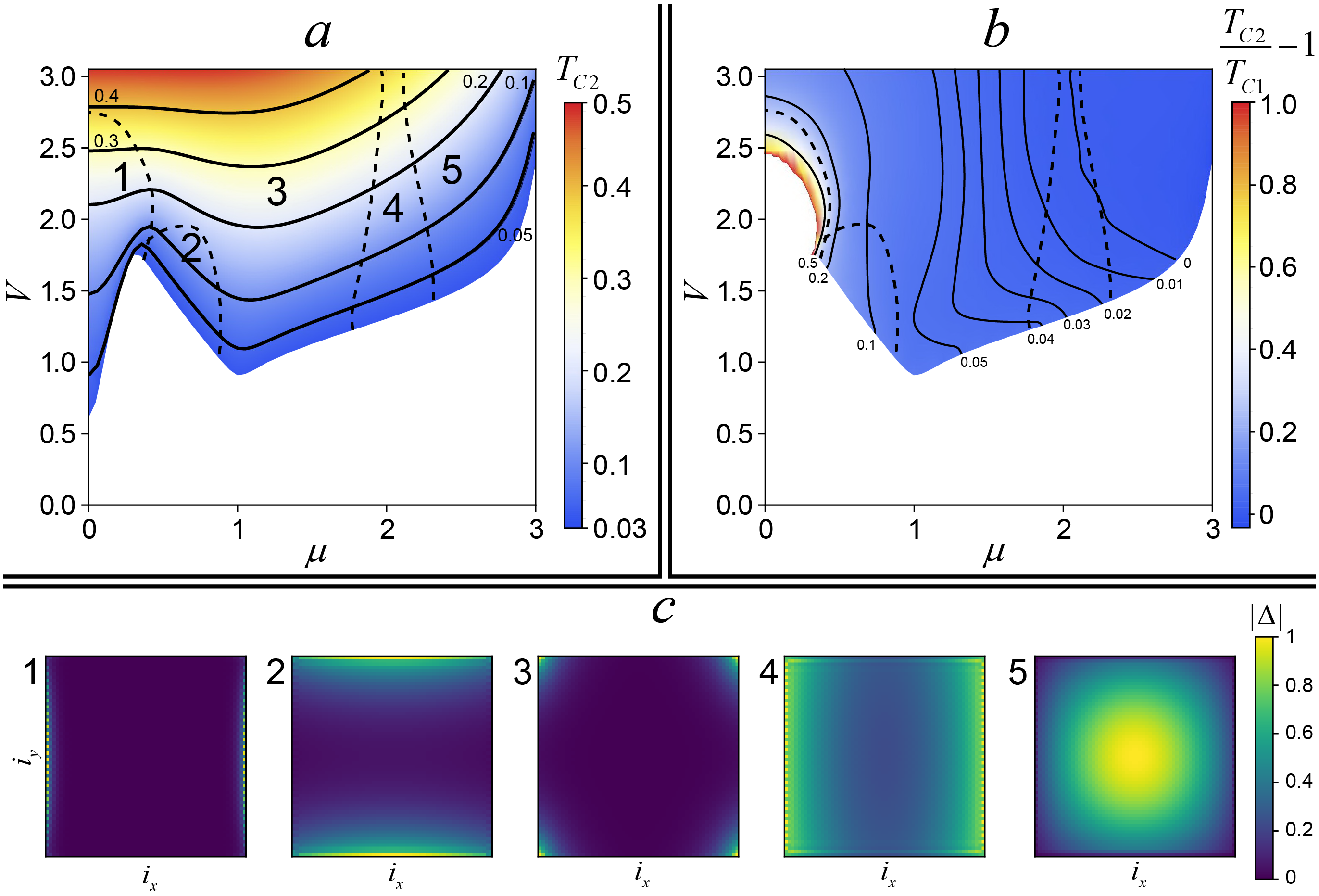}
    \caption{(\textit{a}) Finite rectangular graphene nanoflake ($N_x$ even, $N_y$ odd) phase diagram.
    Solid lines correspond to constant critical temperature curves with $T_c$ written close to the line.
    Big numbers enumerate the regions with different order parameter distributions illustrated in the part (\textit{c}).
    Dashed lines are the 'transition lines' between the regions ($\Delta$ in the corresponding locations are equal).
    (\textit{b}) Relative change in the $T_c$ for the rectangular nanoflake in comparison to the infinite graphene sheet.
    Solid lines are constant-level curves.
    Dashed lines are the same as in the part (\textit{a}).}
    \label{fig: finite 60x61}
\end{figure*}

On the phase diagram (Fig. \ref{fig: finite 60x61}\textit{a}) one can distinguish five regions with different order parameter distributions.
There are four locations where we determine the gap: in the center, in the corners, on vertical and horizontal boundaries.
The gap is the same in all corners due to the system symmetry.
We will use the average gap value for the boundaries because the order parameter oscillates (without sign change) on vertical boundaries and is also not uniform on horizontal ones (it changes close to the corners).
We define regions 1 and 4 as regions where the gap on the vertical boundary is bigger than the gap in three other locations.
In the same way, we define regions 2 (the biggest gap is on the horizontal boundaries), 3 (in the corners), and 5 (in the center).
The order parameter is enhanced on the zigzag edges and normalized $\Delta$ is smaller than 0.001 in the bulk of the sample in the first region (Fig. \ref{fig: finite 60x61}\textit{c}).
In the second region horizontal (armchair) boundaries give rise to the gap enhancement (Fig. \ref{fig: finite 60x61}\textit{c}).
Here gap in the bulk is still small, but the boundaries are only slightly suppressed in the corners.
The biggest region in $V$, $\mu$ parameter space is the third one, where the gap is localized in the corners (Fig. \ref{fig: finite 60x61}\textit{c}).
It is a new gap distribution state which was not observed in nanotubes (Sec. \ref{section: Graphene nanotubes}) because they do not have corners.
Region 4 has an increase in the gap on zigzag edges, however gap in the bulk is also significant (Fig. \ref{fig: finite 60x61}\textit{c}).
In region 5, the gap is suppressed near all boundaries (Fig. \ref{fig: finite 60x61}\textit{c}).

Relative change in the critical temperature in comparison to the infinite graphene sheet is shown in Fig. \ref{fig: finite 60x61}\textit{b} (we still restrict the maximal value to 1).
Note the monotonic decrease of the relative change when increasing band filling ($\mu$).
The biggest increase is still located for small $\mu$ and $V < 2.5$ where the gap is localized on zigzag edges.
In region 2, an increase in $T_c$ has an order of $10 \%$ where the gap is localized on armchair edges.
In region 3, it varies from no gain to $30 \%$ increase.
In region 4, increase is a few percent where bulk comes into play.
Almost the whole of region 5 has a reduction of $T_c$ due to suppression on the boundaries.

\subsubsection{The 'non-closed structure' case}

We deal with three structures illustrated in Fig. \ref{fig: finite structures} (except the top left one) in the subsection.
They have the following common things: two corners are usual ones (like in the previous subsection) and the rest two have an atom with only one bond.
The three structures have different arrangements of the corners, however, their phase diagrams almost coincide.
That is why we will discuss only one geometry: even $N_x$ and $N_y$ case.

\begin{figure*}
    \centering
    \includegraphics[width=0.8\linewidth]{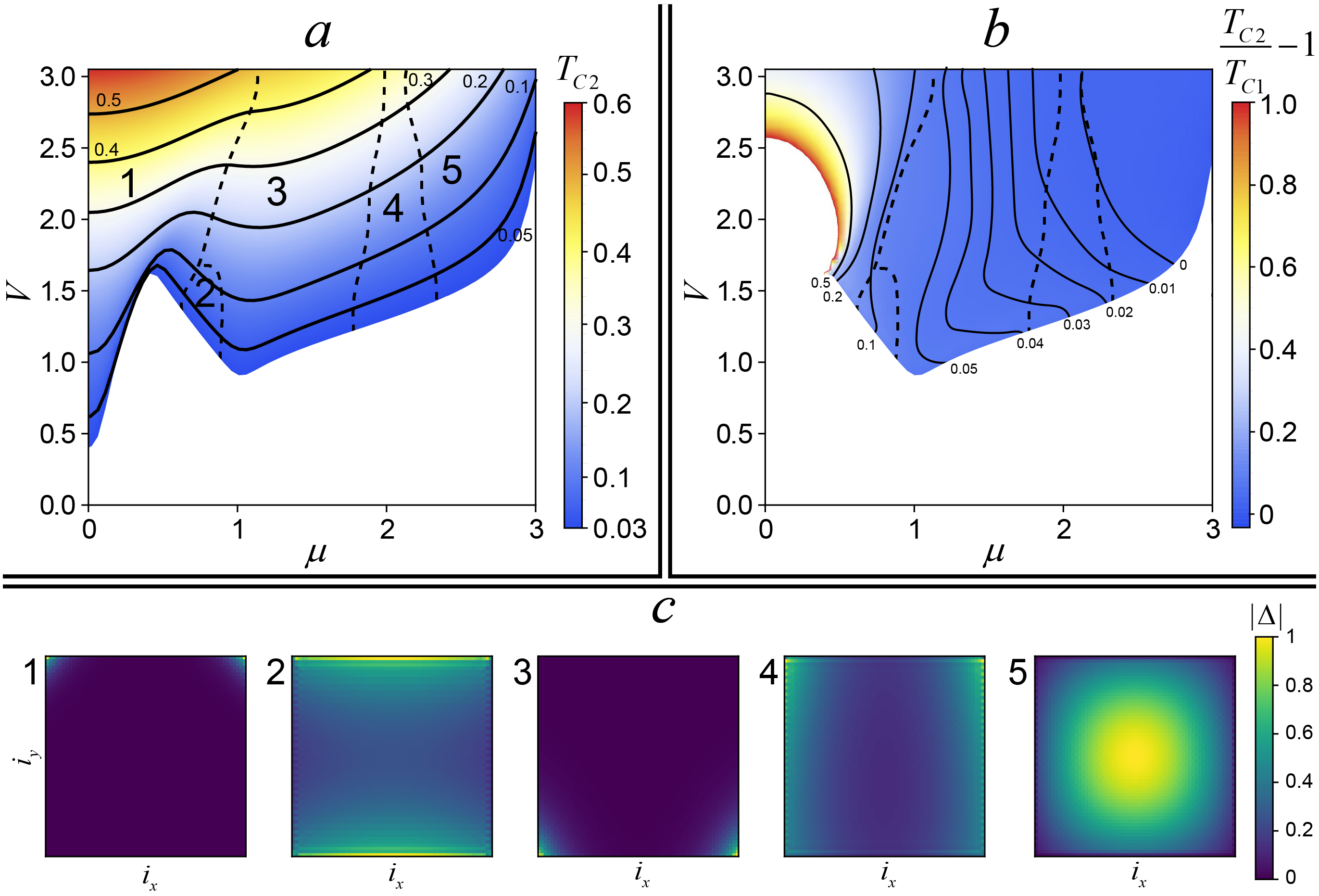}
    \caption{(\textit{a}) Finite rectangular graphene nanoflake ($N_x$ and $N_y$ even) phase diagram.
    Solid lines correspond to constant critical temperature curves with $T_c$ written close to the line.
    Big numbers enumerate the regions with different order parameter distributions illustrated in the part (\textit{c}).
    Dashed lines are the 'transition lines' between the regions ($\Delta$ in the corresponding locations are equal).
    (\textit{b}) Relative change in the $T_c$ for the rectangular nanoflake in comparison to the infinite graphene sheet.
    Solid lines are constant-level curves.
    Dashed lines are the same as in the part (\textit{a}).}
    \label{fig: finite 60x60}
\end{figure*}

We also defined five regions on the phase diagram (Fig. \ref{fig: finite 60x60}\textit{a}) .
Definitions of regions 2, 4, and 5 remain the same: the biggest gap on armchair (horizontal) boundaries, zigzag (vertical) boundaries, and in the center respectively.
However, now we have two different types of corner states: usual corners and corners with a single bond atom.
The gap in the latter type of corner is the biggest in the system in region 1.
The largest gap in the system is located in the usual corners in region 3 of the phase diagram in Fig. \ref{fig: finite 60x60}\textit{a}.
Region 2 in parameter space became smaller in comparison to the 'closed structure' case (Fig. \ref{fig: finite 60x61}\textit{a}) due to the expansion of region 1.
Regions 4 and 5 remained approximately the same.
Note, that maximal critical temperature increased from 0.5 (for nanoribbons and 'closed structure' finite sample) to 0.6 in the same considered range of $\mu$ and $V$.

Gap distributions 2--5 for the nanoflake (Fig. \ref{fig: finite 60x60}\textit{c}) are similar to the described in the previous subsection (Fig. \ref{fig: finite 60x61}\textit{c}).
The gap distribution in region 1  (Fig. \ref{fig: finite 60x61}\textit{c}) is similar to the distribution in region 3.
However, it is localized even in the smaller sample region.
The main reason is an atom with one bond can have incredibly high $T_c$ and due to the proximity effect, it opens a gap for a few neighbouring sites.
It can be considered similar to a single impurity effect.

Relative change in $T_c$  for the structure is shown in Fig. \ref{fig: finite 60x60}\textit{b}.
Here region of $|\mu| > 1$ is similar to the one in Fig. \ref{fig: finite 60x61}\textit{b} so we will discuss only $|\mu| \leq 1$.
The range of $\mu$, $V$ parameters with a relative increase higher than 1 is the biggest in comparison to all considered structures.
For $V$ in the range [1.5; 2] increasing chemical potential leads to a rapid decrease in the relative change in $T_c$ in region 1.

One can explain transitions in the finite sample between regions with different gap distributions from an energetic point of view. From sections \ref{section: infinite} and \ref{section: Graphene nanotubes}, we know critical temperatures for bulk state and boundary (armchair and zigzag) states respectively. Consequently one can calculate boundaries between the three regions (bulk and two boundary states) on $V(\mu)$ phase diagram. The method is described in Appendix \ref{app:comparison} with results quantitatively similar to phase diagrams in Figs. \ref{fig: finite 60x61}\textit{a} and \ref{fig: finite 60x60}\textit{a}.

\section{Conclusions}

In conclusion, recently the problem of superconductivity near boundaries of a BCS superconductor was revisited showing that scattering from the surface is very important and one cannot apply simple approximations for averaging over Friedel oscillations of density of states \cite{samoilenka2020boundary,samoilenka2020pair,benfenati2021boundary,barkman2022elevated,samoilenka2021microscopic}. These references studied the problem in continuum and on a square lattice. Here we studied interplay of this physics with the physics of nontrivial single-electron boundary 
states. To that end, we considered one of the simplest examples: the problem of superconductivity on a honeycomb lattice with s-wave pairing interaction. We found that the boundary superconductivity in that case allows a great diversity of patterns. The gap patterns include surface superconductivity, including the one with normal bulk, and corner superconductivity but also suppression of superconducting gaps at various surfaces. 

For the cases of an armchair and zigzag nanotubes,
there are two possible gap states: enhanced or suppressed gap at the boundary.
The latter state is usually observed for an almost filled (empty) band.
However, for a zigzag nanotube, the such state also exists for a filling close to the M point in the Brillouin zone ($\mu = 1$) and pairing potential $V < 2$.
In the case of an armchair nanotube gap does not depend on the azimuth, however, a zigzag nanotube has nonuniform gap distribution in the azimuth direction due to the alternation of atoms with two and three bonds on the edges.
A zigzag nanotube has a drastically different superconductivity phase diagram (in $V$ and $\mu$ axes) from an infinite sample: in the region of small doping ($|\mu| < 0.4$) pairing potential is much smaller than $V_{\infty}$ for given critical temperature.
In the case of fixing $V$, it means that we can get hundreds of times higher $T_c$ for zigzag nanotube boundaries (because of logarithmic dependence $V$ on $T_c$ for infinite sample).

A finite rectangular honeycomb sample has at least four different gap states.
The first two of them are the boundary states with gap enhancement on the boundaries that were found in nanotubes: either zigzag edge state or armchair edge state.
The third one is a corner state with gap enhancement.
For one out of four rectangular geometries, the state is single, because all corners are identical.
However, three other rectangular geometries have two types of corners: where boundary atoms have two bonds and a type where in the corner one atom has only one bond.
The latter state nonzero gap is localized in a smaller sample region in comparison to the first type corner state.
The corner state with a single bond atom is more energetically favourable than the zigzag boundary type state for small values of doping ($|\mu|< 0.7$) due to lower $V$ for a given critical temperature.
Consequently, the state has an even higher $T_c$.
The fourth gap state is the state where boundaries and corners lead to suppression of $T_c$, which emerges for an almost filled band.

If one considers the superconducting transitions of a half-filled rectangular honeycomb lattice sheet, one will see the following picture during the cooling process. First, local superconductivity emerges in the corners with a single bond atom (if it exists in the sample).
Then nonzero gap appears on zigzag boundaries, and later on armchair boundaries.
Arising of the bulk state depends on the pairing potential: if $V<2t$ critical temperature for graphene should be less than 0.01 K, which is complicated to achieve.
Therefore, one can see the corner and boundary gap states without any bulk superconductivity.

We note that the calculations are based on mean-field approximation and in practice these critical temperatures will be suppressed by fluctuations, however, there are many cases of observation of superconductivity even in zero-dimensional systems. The broader implication of the findings is that they illustrate that the system with normal bulk and nontrivial single-electron surface states can have a strong dependence on critical temperature and gap value on the surface. Some of these features should also persist in multi-layer or twisted bilayer graphene that may also under certain condition exhibit superconductivity only on boundary layers. 

\begin{acknowledgments}
This work was supported by
the Knut and Alice Wallenberg Foundation via the Wallenberg Center for Quantum Technology (WACQT)
and Swedish Research Council Grants 2016-06122, 2018-03659.
We thank Mats Barkman for useful discussions.
\end{acknowledgments}

\clearpage

\appendix

\section{Explicit forms of $H_0$ and $\Delta$ matrices and derivation of thermal averages} \label{app:ap1}
Using coordinate form of vectors $\textbf{i} = (n,m)$ and $\textbf{j} = (p,r)$ one can get explicit expression for matrices $H_0 (\textbf{i},\textbf{j})$ and $\Delta (\textbf{i},\textbf{j})$:
\begin{widetext}
  \begin{equation}
        H_0 (\textbf{i},\textbf{j}) = -
    \begin{pmatrix}
        \mu \delta_{\textbf{i},\textbf{j}} &&  \delta_{\textbf{i},\textbf{j}} + \delta_{p,n-1} (\delta_{r,m + 1} + \delta_{r,m - 1}) \\
         \delta_{\textbf{i},\textbf{j}} + \delta_{p,n+1} (\delta_{r,m + 1} + \delta_{r,m - 1}) && \mu \delta_{\textbf{i},\textbf{j}}
    \end{pmatrix},
  \end{equation}  
\end{widetext}
\begin{equation}
    \Delta (\textbf{i},\textbf{j}) = \begin{pmatrix}
        \delta_{\textbf{i},\textbf{j}} \Delta_{\textbf{i},A} && 0\\
        0 && \delta_{\textbf{i},\textbf{j}} \Delta_{\textbf{i},B}
    \end{pmatrix}.
\end{equation}
They satisfy relations $H_0^* (\textbf{i},\textbf{j}) = H_0 (\textbf{i},\textbf{j})$, $\Delta ^T (\textbf{i},\textbf{j}) = \Delta (\textbf{i},\textbf{j})$ as a consequence half of energies $E^\nu$ have to be positive. This comes from the theorem that if we know eigenvector $(u^{\nu}_\textbf{i} , y^{\nu}_\textbf{i} , v^{\nu}_\textbf{i} , z^{\nu}_\textbf{i})^T$ corresponding to eigenenergy $E^\nu$ for Eq.(\ref{eq:BdG_eq}), then eigenvector $(-v^{\nu}_\textbf{i} , -z^{\nu}_\textbf{i} , u^{\nu}_\textbf{i} , y^{\nu}_\textbf{i})\da$ is the solution for the same equations with eigenvalue $-E^\nu$.

We need to find thermal average $\langle a_{\textbf{i},\downarrow} a_{\textbf{i},\uparrow} \rangle$ ($\langle b_{\textbf{i},\downarrow} b_{\textbf{i},\uparrow} \rangle$ can be written by analogy) to calculate order parameter Eq. (\ref{eq:delta_definition}). Let's substitute definition of $a_{\textbf{i},\sigma}$ from Eq. (\ref{eq:Bogoliubov_transformation}) and simplify expression like in \cite{de1966superconductivity}
\begin{equation}
    \begin{gathered}
        \langle a_{\textbf{i},\downarrow} a_{\textbf{i},\uparrow} \rangle = 
        \langle \sum_\nu^{'} \sum_\lambda^{'} (u_{\textbf{i}}^\nu \gamma_{\nu, \downarrow} + v_{\textbf{i}}^{\nu *} \gamma_{\nu, \uparrow}\da)
        (u_{\textbf{i}}^\lambda \gamma_{\lambda, \uparrow} - v_{\textbf{i}}^{\lambda *} \gamma_{\lambda, \downarrow}\da) \rangle \\
        = \sum_\nu^{'} \sum_\lambda^{'}
            (
             u_{\textbf{i}}^\nu u_{\textbf{i}}^\lambda \langle \gamma_{\nu, \downarrow} \gamma_{\lambda, \uparrow} \rangle
             - v_{\textbf{i}}^{\nu *} v_{\textbf{i}}^{\lambda *} \langle \gamma_{\nu, \uparrow}\da \gamma_{\lambda, \downarrow}\da \rangle \\
             + v_{\textbf{i}}^{\nu *} u_{\textbf{i}}^\lambda \langle \gamma_{\nu, \uparrow}\da \gamma_{\lambda, \uparrow} \rangle
             - u_{\textbf{i}}^\nu v_{\textbf{i}}^{\lambda *} \langle \gamma_{\nu, \downarrow} \gamma_{\lambda, \downarrow}\da \rangle
            ) \\
        = \sum_\nu^{'} \sum_\lambda^{'}
            (
             0
             - 0
             + v_{\textbf{i}}^{\nu *} u_{\textbf{i}}^\lambda \delta_{\nu,\lambda} f(E^\nu)
             - u_{\textbf{i}}^\nu v_{\textbf{i}}^{\lambda *} \delta_{\nu,\lambda} f(-E^\nu)
            ) \\
        = \sum_\nu^{'} u_{\textbf{i}}^\nu v_{\textbf{i}}^{\nu *}
            (f(E^\nu) -  f(-E^\nu)) = - \sum_\nu^{'} u_{\textbf{i}}^\nu v_{\textbf{i}}^{\nu *} \tanh{\frac{E^\nu}{2 T}},
    \end{gathered}
\end{equation}
\begin{equation}
\begin{gathered}
        \langle b_{\textbf{i},\downarrow} b_{\textbf{i},\uparrow} \rangle = 
        \langle \sum_\nu^{'} \sum_\lambda^{'} (y_{\textbf{i}}^\nu \gamma_{\nu, \downarrow} + z_{\textbf{i}}^{\nu *} \gamma_{\nu, \uparrow}\da)
        (y_{\textbf{i}}^\lambda \gamma_{\lambda, \uparrow} - z_{\textbf{i}}^{\lambda *} \gamma_{\lambda, \downarrow}\da) \rangle \\
        = - \sum_\nu^{'} y_{\textbf{i}}^\nu z_{\textbf{i}}^{\nu *} \tanh{\frac{E^\nu}{2 T}},
\end{gathered}
\end{equation}
where $f(E)$ -- the Fermi distribution function ($f(E) = (1+e^{E/{T}})^{-1}$). Here we used following definitions for thermal average: $\langle \gamma_{\alpha,\sigma1}\da \gamma_{\beta,\sigma2} \rangle = \delta_{\alpha, \beta} \delta_{\sigma1,\sigma2} f(E^\alpha)$, $\langle \gamma_{\alpha,\sigma1} \gamma_{\beta,\sigma2}\da \rangle = \delta_{\alpha, \beta} \delta_{\sigma1,\sigma2} f(-E^\alpha)$, $\langle \gamma_{\alpha,\sigma1} \gamma_{\beta,\sigma2} \rangle = \langle \gamma_{\alpha,\sigma1}\da \gamma_{\beta,\sigma2}\da \rangle = 0$. The self-consistency condition for the energy gap (Eq. (\ref{eq:delta_definition})) now has form (\ref{eq:delta_in_vectors}).

\section{Derivation of self-consistent conditions for infinite case} \label{app:ap2}
Eigenvectors that correspond to eigenvalues Eq. (\ref{eq:energies with delta}) have a form
\begin{equation}
\label{eq:eigenvectors}
    \mqty(\mathcal{U}_{\textbf{k},s} \\ \mathcal{Y}_{\textbf{k},s} \\ \mathcal{V}_{\textbf{k},s} \\ \mathcal{Z}_{\textbf{k},s}) =
    \medmath{\frac{1}{2\sqrt{E_s (\epsilon_s + E_s)}} }
    \mqty(- \frac{s(1 + 2 e^{-i k_x} \cos{k_y})(\epsilon_s + E_s)}{\epsilon_0 (k_x, k_y)}
    \\
          \epsilon_s + E_s
    \\
          - \frac{s \Delta^* (1 + 2 e^{-i k_x} \cos{k_y})}{\epsilon_0 (k_x, k_y)}
    \\
          \Delta^*).
\end{equation}
These eigenvectors are normalized with the rule $(\mathcal{U}_{\textbf{k},s}, \mathcal{Y}_{\textbf{k},s}, \mathcal{V}_{\textbf{k},s}, \mathcal{Z}_{\textbf{k},s}) (\mathcal{U}_{\textbf{k},s}, \mathcal{Y}_{\textbf{k},s}, \mathcal{V}_{\textbf{k},s}, \mathcal{Z}_{\textbf{k},s})\da = 1$.

Let's simplify expression for order parameter (Eq. (\ref{eq:delta_in_vectors})) using results (\ref{eq:Bloch_theorem}) and (\ref{eq:eigenvectors}):
\begin{equation}
\label{eq:delta_almost_final}
\begin{gathered}
    \Delta_{\textbf{i},B} = \Delta = V \frac{1}{N_x N_y / 2} \sum_{s = \pm 1} \sum_{k_x, k_y}
      \mathcal{Y}_\textbf{k} \mathcal{Z}^{*}_\textbf{k} \tanh{\frac{E_s}{2 T}} 
     \\
     = \frac{2V}{N_x N_y} \sum_{s = \pm 1} \sum_{k_x, k_y} \frac{(\epsilon_s + E_s) \Delta }{4 E_s (\epsilon_s + E_s)}  \tanh{\frac{E_s}{2 T}}
     \\
     = \frac{V}{2 N_x N_y} \sum_{s = \pm 1} \sum_{k_x, k_y} \frac{\Delta}{E_s} \tanh{\frac{E_s}{2 T}}.
\end{gathered}
\end{equation}
Here we switched from summation over $\nu$ in (\ref{eq:delta_in_vectors}) to summation over $s$ only for positive energies $E_s$ (\ref{eq:energies with delta}). This expression can be further simplified in assumption of constant $\Delta$ and with change summation over $k_x$ and $k_y$ to integration in the limit $N_x, N_y \rightarrow \infty$:
\begin{equation}
\label{eq:delta_uniform}
\begin{gathered}
    1 = \lim_{N_x, N_y \to \infty} \frac{V}{2 N_x N_y} \sum_{s = \pm 1} \sum_{k_x, k_y} \frac{\tanh{\frac{E_s}{2 T}}}{E_s}
    \\
      = \lim_{N_x, N_y \to \infty} \medmath{\frac{V}{2 N_x N_y} \sum_{s = \pm 1} \frac{N_x N_y}{2 S_{\text{1st BZ}}} \iint_{\text{1st BZ}} dk_x dk_y \frac{\tanh{\frac{E_s}{2 T}}}{E_s}}
    \\
      = \frac{V}{4 S_{\text{1st BZ}}} \sum_{s = \pm 1} \iint_{\text{1st BZ}} dk_x dk_y \frac{\tanh{\frac{E_s}{2 T}}}{E_s}. \\
\mycomment{      = \frac{V}{4 S_{\text{1st BZ}}} \iint_{\text{1st BZ}} dk_x dk_y \Biggl( \frac{\tanh{\frac{\sqrt{\epsilon^{2}_+ + \Delta\Delta^*}}{2 T}}}{\sqrt{\epsilon^{2}_+ + \Delta\Delta^*}}
    \\
      + \frac{\tanh{\frac{\sqrt{\epsilon^{2}_- + \Delta\Delta^*}}{2 T}}}{\sqrt{\epsilon^{2}_- + \Delta\Delta^*}} \Biggr).
}
\end{gathered}
\end{equation}

Considering self-consistent relation for $A$ sites one can come to an identical result.

\section{Zero temperature limit of self-consistent equation} \label{app:ap3}
Let us look at self-consistent equation for a general 2d lattice case:
\begin{equation} \label{eq:int1}
    \frac{1}{V} = C \iint_{\text{1st BZ}} dk_x dk_y \frac{\tanh{\frac{E}{2 T}}}{E},
\end{equation}
where $C$ is some coefficient proportional to the area of the 1st BZ, $E$ is energy which includes shift by chemical potential.
We can divide the integral into two parts depending on the energy value: energies above and below some threshold ($E_\text{tr}$). We can clarify constraints to the threshold in the form $T \ll E_\text{tr} \ll 1$. Remind that we are approaching zero temperature, so constraints can be satisfied. Equation (\ref{eq:int1}) can be rewritten as
\begin{equation} \label{eq:int2}
\begin{split}
    \frac{1}{V} =& C \iint_{|E|>E_\text{tr}} \frac{dk_x dk_y }{|E|} \\
    &+ C \iint_{|E| \leq E_\text{tr}} dk_x dk_y \frac{\tanh{\frac{E}{2 T}}}{E}.
\end{split}
\end{equation}
The first integral does not depend on the temperature. Further, we look only at the second integral. Energy is small, so it can be expanded into series:
\begin{equation} \label{eq:expansion}
    E \approx \alpha(\mu, k_\parallel) \cdot k_\perp,
\end{equation}
where we chose another momentum coordinates: $k_\parallel$ is the momentum parallel to the Fermi surface and $k_\perp$ is the momentum perpendicular to the above mentioned direction; $\alpha$ is the modulus of gradient in the point $(\mu, k_\parallel)$. The second integral in Eq. (\ref{eq:int2}) can be calculated as follows
\begin{equation} \label{eq:estimation1}
\begin{gathered}
    \iint_{|E| \leq E_\text{tr}} dk_x dk_y \frac{\tanh{\frac{E}{2 T}}}{E} \\
    \approx \iint_{|E| \leq E_\text{tr}} dk_\parallel dk_\perp \frac{\tanh{\frac{E}{2 T}}}{E} = \frac{2 l_{k_\parallel}}{\alpha} \int_{0}^{E_\text{tr}} dE \frac{\tanh{\frac{E}{2 T}}}{E} \\
    \approx \frac{2 l_{k_\parallel}}{\alpha} \left( \ln{\frac{E_\text{tr}}{2 T}} + \ln{\frac{4 e^\gamma}{\pi}} \right),
\end{gathered}
\end{equation}
where $l_{k_\parallel}$ is the length of Fermi surface, $\gamma$ is Euler's constant. Here we assumed a constant modulus of the gradient. However, it depends on $k_\parallel$ in general. Result (\ref{eq:estimation1}) can be used as a lower boundary for the integral if we take $\alpha_\text{max}$ for a given chemical potential and vice versa.

Final result for self-consistent equation (\ref{eq:int1}) is following
\begin{equation}
    \frac{1}{V} = C \left( \iint_{|E|>E_\text{tr}} \frac{dk_x dk_y }{|E|} + \frac{2 l_{k_\parallel}}{\alpha} \ln{\frac{4 e^\gamma}{\pi}} + \frac{2 l_{k_\parallel}}{\alpha} \ln{\frac{E_\text{tr}}{2 T}}\right).
\end{equation}
The first two terms do not depend on temperature, so when approaching absolute zero one can neglect them in comparison to the last one.

The same result can be obtained in a bit different way.
Let us calculate partial derivative of Eq. (\ref{eq:int1}) with respect to the temperature:
\begin{equation} \label{eq:partial T}
    \frac{\partial \left( \frac{1}{V} \right)}{\partial T} = - \frac{C}{2 T} \iint_{\text{1st BZ}} \frac{dk_x dk_y}{T \cosh^{2} \left( \frac{E}{2 T} \right)}.
\end{equation}
The integral has the following bounds
\begin{equation} \label{eq:bounds}
\begin{gathered}
    I_\text{min} = \iint_{\text{1st BZ}} \frac{dk_x dk_y}{T \exp \left( \frac{|E|}{T} \right)}, \\
    I_\text{max} = 4 I_\text{min}.
\end{gathered}
\end{equation}
The exponent is localized in the region $|E| \lesssim T$. Using expansion Eq. (\ref{eq:expansion}) and switching to coordinates $k_\parallel$, $k_\perp$ bounds Eq. (\ref{eq:bounds}) have form
\begin{equation} \label{eq:bounds final}
    I_\text{min} \approx \frac{2 l_{k_\parallel}}{\alpha_\text{max}}, \qquad
    I_\text{max} \approx \frac{8 l_{k_\parallel}}{\alpha_\text{min}}.
\end{equation}
Therefore, partial derivative with respect to $T$ Eq. (\ref{eq:partial T}) has bounds
\begin{equation}
    - \frac{l_{k_\parallel} C}{\alpha_\text{max} T} \lesssim \frac{\partial \left( \frac{1}{V} \right)}{\partial T} \lesssim - \frac{4 l_{k_\parallel} C}{\alpha_\text{min} T}.
\end{equation}

The result has the same consequence: we have a divergence of the partial derivative when approaching absolute zero which means that $V \rightarrow 0$.
Note, that the derivation works only in the case when Fermi surface has nonzero length.
The same conclusion can be obtained also for multiband systems.

\section{Comparison of the systems} \label{app:comparison}

In the Sec. \ref{section: Graphene nanotubes} we found configurations for nanotubes (similar configurations have infinite nanoribbons).
They correspond to boundary states in finite rectangular nanoflakes.
In the finite sample, one can have the following configurations: bulk state, corner states, or two types of boundary states.
Usually, gap distribution is a superposition of the above-mentioned states, but in the majority of cases, one of the states clearly dominates.
The system chooses a configuration with the lowest energy.
In reality, when $\mu$ and $V$ are fixed it is a configuration with the highest possible $T_c$.
In further discussion, we fix $T_c$ and plot $V(\mu)$, so the most favourable configuration has the lowest $V$.

\begin{figure}
    \centering
    \includegraphics[width=0.99\linewidth]{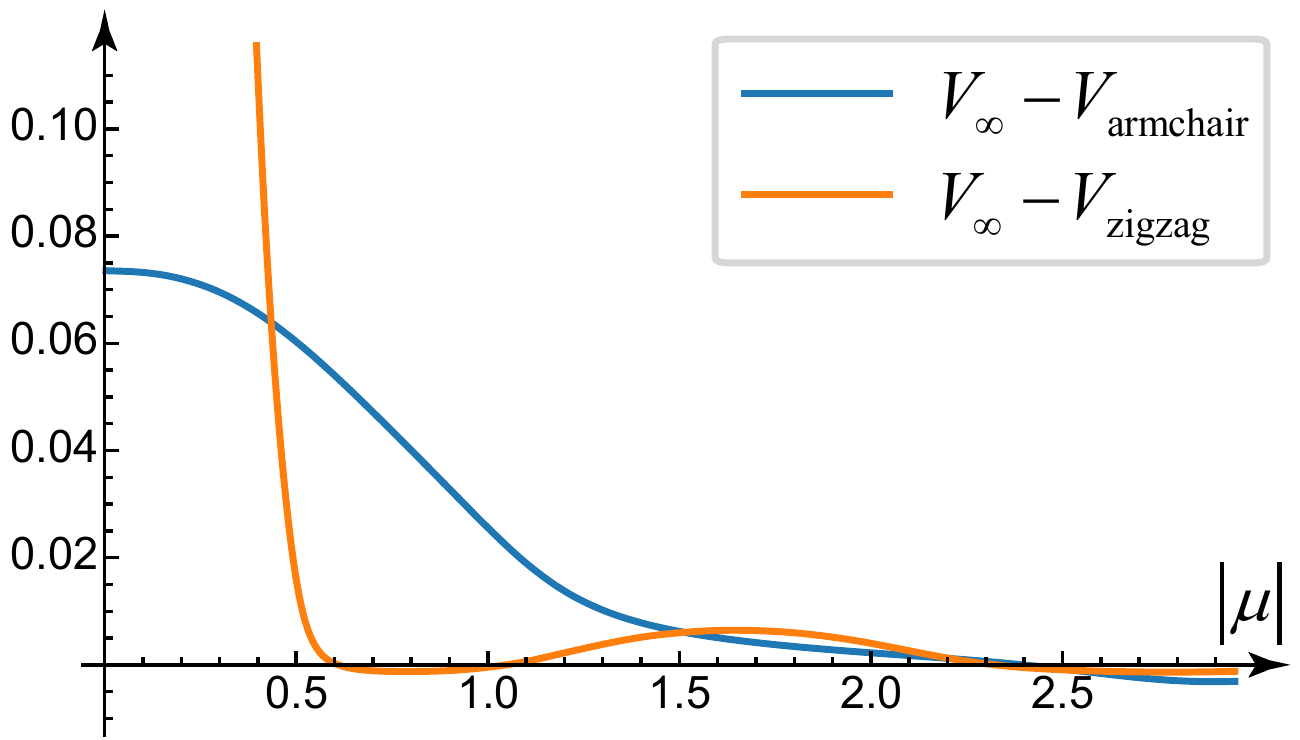}
    \caption{The difference between pairing potential for the infinite system and the value for nanotubes (armchair and zigzag) as a function of chemical potential for $T_c = 0.1$.}
    \label{fig: comparison}
\end{figure}

Figure \ref{fig: comparison} shows the difference between pairing potential for the infinite system and the value for nanotubes from Sec. \ref{section: Graphene nanotubes} as a function of chemical potential.
A finite system chooses a state with the lowest $V$, hence the plotted value should be the biggest among positive ones (in the case we have one of the boundary states) or if they are negative system chooses a bulk state.
In the regions $\mu \in [0; 0.44) \cup (1.50; 2.23)$ $V$ in the zigzag nanotube is the smallest, that is why finite nanoflake chooses to have a gap on zigzag edges.
In the regions $\mu \in (0.44; 1.50) \cup (2.23; 2.37)$ $V$ in the armchair nanotube is the smallest, which is why the finite system prefers to have a gap on the armchair edges.
In the region $\mu \in (2.37; 3)$ both plots are below zero.
It means that it is preferable to have a bulk state.
It is quantitatively consistent with the phase diagram in Fig. \ref{fig: finite 60x61}\textit{a} (for the same $T_c = 0.1$): for $|\mu| < 0.4$ 'closed structure' system prefers zigzag edge states, then for $|\mu| \in (0.48; 0.85)$ armchair edge states.
In the region of $|\mu| \in (1.85; 2.28)$ again zigzag edges are favourable and for $|\mu| \in (2.28; 3]$ bulk state dominates.
Our discussion in the appendix does not take into account corner states which are for sure important in finite samples.
To find out the phase diagram corresponding only to the corner states one should investigate a semi-infinite rectangular corner system.

When applying the results to a 'non-closed structure' finite sample (Fig. \ref{fig: finite 60x60}\textit{a}) we have a smaller region of chemical potential where the boundary states favourable.
It is due to the existence of two types of corner states.
In the system, the approach describes the transition between regions 4 and 5 which is located at $\mu = 2.28$ (the appendix approach predicts $\mu = 2.4$ at $T_c = 0.1$).

\bibliography{references.bib}

\begin{thebibliography}{42}%
\makeatletter
\providecommand \@ifxundefined [1]{%
 \@ifx{#1\undefined}
}%
\providecommand \@ifnum [1]{%
 \ifnum #1\expandafter \@firstoftwo
 \else \expandafter \@secondoftwo
 \fi
}%
\providecommand \@ifx [1]{%
 \ifx #1\expandafter \@firstoftwo
 \else \expandafter \@secondoftwo
 \fi
}%
\providecommand \natexlab [1]{#1}%
\providecommand \enquote  [1]{``#1''}%
\providecommand \bibnamefont  [1]{#1}%
\providecommand \bibfnamefont [1]{#1}%
\providecommand \citenamefont [1]{#1}%
\providecommand \href@noop [0]{\@secondoftwo}%
\providecommand \href [0]{\begingroup \@sanitize@url \@href}%
\providecommand \@href[1]{\@@startlink{#1}\@@href}%
\providecommand \@@href[1]{\endgroup#1\@@endlink}%
\providecommand \@sanitize@url [0]{\catcode `\\12\catcode `\$12\catcode
  `\&12\catcode `\#12\catcode `\^12\catcode `\_12\catcode `\%12\relax}%
\providecommand \@@startlink[1]{}%
\providecommand \@@endlink[0]{}%
\providecommand \url  [0]{\begingroup\@sanitize@url \@url }%
\providecommand \@url [1]{\endgroup\@href {#1}{\urlprefix }}%
\providecommand \urlprefix  [0]{URL }%
\providecommand \Eprint [0]{\href }%
\providecommand \doibase [0]{https://doi.org/}%
\providecommand \selectlanguage [0]{\@gobble}%
\providecommand \bibinfo  [0]{\@secondoftwo}%
\providecommand \bibfield  [0]{\@secondoftwo}%
\providecommand \translation [1]{[#1]}%
\providecommand \BibitemOpen [0]{}%
\providecommand \bibitemStop [0]{}%
\providecommand \bibitemNoStop [0]{.\EOS\space}%
\providecommand \EOS [0]{\spacefactor3000\relax}%
\providecommand \BibitemShut  [1]{\csname bibitem#1\endcsname}%
\let\auto@bib@innerbib\@empty
%</preamble>
\bibitem [{\citenamefont {de~Gennes}(1964)}]{de1964boundary}%
  \BibitemOpen
  \bibfield  {author} {\bibinfo {author} {\bibfnamefont {P.}~\bibnamefont
  {de~Gennes}},\ }\bibfield  {title} {\bibinfo {title} {Boundary effects in
  superconductors},\ }\href@noop {} {\bibfield  {journal} {\bibinfo  {journal}
  {Reviews of Modern Physics}\ }\textbf {\bibinfo {volume} {36}},\ \bibinfo
  {pages} {225} (\bibinfo {year} {1964})}\BibitemShut {NoStop}%
\bibitem [{\citenamefont {de~Gennes}(1966)}]{de1966superconductivity}%
  \BibitemOpen
  \bibfield  {author} {\bibinfo {author} {\bibfnamefont {P.}~\bibnamefont
  {de~Gennes}},\ }\href@noop {} {\emph {\bibinfo {title} {Superconductivity of
  metals and alloys}}}\ (\bibinfo  {publisher} {WA Benjamin, Inc., New York},\
  \bibinfo {year} {1966})\BibitemShut {NoStop}%
\bibitem [{\citenamefont {Caroli}\ \emph {et~al.}(1963)\citenamefont {Caroli},
  \citenamefont {De~Gennes},\ and\ \citenamefont {Matricon}}]{CdGM_Coherence}%
  \BibitemOpen
  \bibfield  {author} {\bibinfo {author} {\bibfnamefont {C.}~\bibnamefont
  {Caroli}}, \bibinfo {author} {\bibfnamefont {P.}~\bibnamefont {De~Gennes}},\
  and\ \bibinfo {author} {\bibfnamefont {J.}~\bibnamefont {Matricon}},\
  }\bibfield  {title} {\bibinfo {title} {Coherence length and penetration depth
  of dirty superconductors},\ }\href@noop {} {\bibfield  {journal} {\bibinfo
  {journal} {Physik der kondensierten Materie}\ }\textbf {\bibinfo {volume}
  {1}},\ \bibinfo {pages} {176} (\bibinfo {year} {1963})}\BibitemShut {NoStop}%
\bibitem [{\citenamefont {{Caroli, C.}}\ \emph {et~al.}(1962)\citenamefont
  {{Caroli, C.}}, \citenamefont {{De Gennes, P.G.}},\ and\ \citenamefont
  {{Matricon, J.}}}]{CdGM_french}%
  \BibitemOpen
  \bibfield  {author} {\bibinfo {author} {\bibnamefont {{Caroli, C.}}},
  \bibinfo {author} {\bibnamefont {{De Gennes, P.G.}}},\ and\ \bibinfo {author}
  {\bibnamefont {{Matricon, J.}}},\ }\bibfield  {title} {\bibinfo {title} {Sur
  certaines propri\'et\'es des alliages supraconducteurs non magn\'etiques},\
  }\href {https://doi.org/10.1051/jphysrad:019620023010070700} {\bibfield
  {journal} {\bibinfo  {journal} {J. Phys. Radium}\ }\textbf {\bibinfo {volume}
  {23}},\ \bibinfo {pages} {707} (\bibinfo {year} {1962})}\BibitemShut
  {NoStop}%
\bibitem [{\citenamefont {Abrikosov}(1965)}]{abrikosov1965concerning}%
  \BibitemOpen
  \bibfield  {author} {\bibinfo {author} {\bibfnamefont {A.}~\bibnamefont
  {Abrikosov}},\ }\bibfield  {title} {\bibinfo {title} {Concerning surface
  superconductivity in strong magnetic fields},\ }\href@noop {} {\bibfield
  {journal} {\bibinfo  {journal} {Sov. Phys. JETP}\ }\textbf {\bibinfo {volume}
  {20}},\ \bibinfo {pages} {480} (\bibinfo {year} {1965})}\BibitemShut
  {NoStop}%
\bibitem [{\citenamefont {Samoilenka}\ and\ \citenamefont
  {Babaev}(2020)}]{samoilenka2020boundary}%
  \BibitemOpen
  \bibfield  {author} {\bibinfo {author} {\bibfnamefont {A.}~\bibnamefont
  {Samoilenka}}\ and\ \bibinfo {author} {\bibfnamefont {E.}~\bibnamefont
  {Babaev}},\ }\bibfield  {title} {\bibinfo {title} {Boundary states with
  elevated critical temperatures in bardeen-cooper-schrieffer
  superconductors},\ }\href@noop {} {\bibfield  {journal} {\bibinfo  {journal}
  {Physical Review B}\ }\textbf {\bibinfo {volume} {101}},\ \bibinfo {pages}
  {134512} (\bibinfo {year} {2020})}\BibitemShut {NoStop}%
\bibitem [{\citenamefont {Samoilenka}\ \emph {et~al.}(2020)\citenamefont
  {Samoilenka}, \citenamefont {Barkman}, \citenamefont {Benfenati},\ and\
  \citenamefont {Babaev}}]{samoilenka2020pair}%
  \BibitemOpen
  \bibfield  {author} {\bibinfo {author} {\bibfnamefont {A.}~\bibnamefont
  {Samoilenka}}, \bibinfo {author} {\bibfnamefont {M.}~\bibnamefont {Barkman}},
  \bibinfo {author} {\bibfnamefont {A.}~\bibnamefont {Benfenati}},\ and\
  \bibinfo {author} {\bibfnamefont {E.}~\bibnamefont {Babaev}},\ }\bibfield
  {title} {\bibinfo {title} {Pair-density-wave superconductivity of faces,
  edges, and vertices in systems with imbalanced fermions},\ }\href@noop {}
  {\bibfield  {journal} {\bibinfo  {journal} {Physical Review B}\ }\textbf
  {\bibinfo {volume} {101}},\ \bibinfo {pages} {054506} (\bibinfo {year}
  {2020})}\BibitemShut {NoStop}%
\bibitem [{\citenamefont {Benfenati}\ \emph {et~al.}(2021)\citenamefont
  {Benfenati}, \citenamefont {Samoilenka},\ and\ \citenamefont
  {Babaev}}]{benfenati2021boundary}%
  \BibitemOpen
  \bibfield  {author} {\bibinfo {author} {\bibfnamefont {A.}~\bibnamefont
  {Benfenati}}, \bibinfo {author} {\bibfnamefont {A.}~\bibnamefont
  {Samoilenka}},\ and\ \bibinfo {author} {\bibfnamefont {E.}~\bibnamefont
  {Babaev}},\ }\bibfield  {title} {\bibinfo {title} {Boundary effects in
  two-band superconductors},\ }\href@noop {} {\bibfield  {journal} {\bibinfo
  {journal} {Physical Review B}\ }\textbf {\bibinfo {volume} {103}},\ \bibinfo
  {pages} {144512} (\bibinfo {year} {2021})}\BibitemShut {NoStop}%
\bibitem [{\citenamefont {Barkman}\ \emph {et~al.}(2022)\citenamefont
  {Barkman}, \citenamefont {Samoilenka}, \citenamefont {Benfenati},\ and\
  \citenamefont {Babaev}}]{barkman2022elevated}%
  \BibitemOpen
  \bibfield  {author} {\bibinfo {author} {\bibfnamefont {M.}~\bibnamefont
  {Barkman}}, \bibinfo {author} {\bibfnamefont {A.}~\bibnamefont {Samoilenka}},
  \bibinfo {author} {\bibfnamefont {A.}~\bibnamefont {Benfenati}},\ and\
  \bibinfo {author} {\bibfnamefont {E.}~\bibnamefont {Babaev}},\ }\bibfield
  {title} {\bibinfo {title} {Elevated critical temperature at bcs
  superconductor-band insulator interfaces},\ }\href@noop {} {\bibfield
  {journal} {\bibinfo  {journal} {arXiv preprint arXiv:2201.11614}\ } (\bibinfo
  {year} {2022})}\BibitemShut {NoStop}%
\bibitem [{\citenamefont {Samoilenka}\ and\ \citenamefont
  {Babaev}(2021)}]{samoilenka2021microscopic}%
  \BibitemOpen
  \bibfield  {author} {\bibinfo {author} {\bibfnamefont {A.}~\bibnamefont
  {Samoilenka}}\ and\ \bibinfo {author} {\bibfnamefont {E.}~\bibnamefont
  {Babaev}},\ }\bibfield  {title} {\bibinfo {title} {Microscopic derivation of
  superconductor-insulator boundary conditions for ginzburg-landau theory
  revisited: Enhanced superconductivity at boundaries with and without magnetic
  field},\ }\href@noop {} {\bibfield  {journal} {\bibinfo  {journal} {Physical
  Review B}\ }\textbf {\bibinfo {volume} {103}},\ \bibinfo {pages} {224516}
  (\bibinfo {year} {2021})}\BibitemShut {NoStop}%
\bibitem [{\citenamefont {Hainzl}\ \emph {et~al.}(2022)\citenamefont {Hainzl},
  \citenamefont {Roos},\ and\ \citenamefont {Seiringer}}]{hainzl2022boundary}%
  \BibitemOpen
  \bibfield  {author} {\bibinfo {author} {\bibfnamefont {C.}~\bibnamefont
  {Hainzl}}, \bibinfo {author} {\bibfnamefont {B.}~\bibnamefont {Roos}},\ and\
  \bibinfo {author} {\bibfnamefont {R.}~\bibnamefont {Seiringer}},\ }\bibfield
  {title} {\bibinfo {title} {Boundary superconductivity in the bcs model},\
  }\href@noop {} {\bibfield  {journal} {\bibinfo  {journal} {arXiv preprint
  arXiv:2201.08090}\ } (\bibinfo {year} {2022})}\BibitemShut {NoStop}%
\bibitem [{\citenamefont {Fink}\ and\ \citenamefont
  {Joiner}(1969)}]{fink1969surface}%
  \BibitemOpen
  \bibfield  {author} {\bibinfo {author} {\bibfnamefont {H.~J.}\ \bibnamefont
  {Fink}}\ and\ \bibinfo {author} {\bibfnamefont {W.~C.~H.}\ \bibnamefont
  {Joiner}},\ }\bibfield  {title} {\bibinfo {title} {Surface nucleation and
  boundary conditions in superconductors},\ }\href
  {https://doi.org/10.1103/PhysRevLett.23.120} {\bibfield  {journal} {\bibinfo
  {journal} {Phys. Rev. Lett.}\ }\textbf {\bibinfo {volume} {23}},\ \bibinfo
  {pages} {120} (\bibinfo {year} {1969})}\BibitemShut {NoStop}%
\bibitem [{\citenamefont {Lortz}\ \emph {et~al.}(2006)\citenamefont {Lortz},
  \citenamefont {Tomita}, \citenamefont {Wang}, \citenamefont {Junod},
  \citenamefont {Schilling}, \citenamefont {Masui},\ and\ \citenamefont
  {Tajima}}]{lortz2006origin}%
  \BibitemOpen
  \bibfield  {author} {\bibinfo {author} {\bibfnamefont {R.}~\bibnamefont
  {Lortz}}, \bibinfo {author} {\bibfnamefont {T.}~\bibnamefont {Tomita}},
  \bibinfo {author} {\bibfnamefont {Y.}~\bibnamefont {Wang}}, \bibinfo {author}
  {\bibfnamefont {A.}~\bibnamefont {Junod}}, \bibinfo {author} {\bibfnamefont
  {J.}~\bibnamefont {Schilling}}, \bibinfo {author} {\bibfnamefont
  {T.}~\bibnamefont {Masui}},\ and\ \bibinfo {author} {\bibfnamefont
  {S.}~\bibnamefont {Tajima}},\ }\bibfield  {title} {\bibinfo {title} {On the
  origin of the double superconducting transition in overdoped yba2cu3ox},\
  }\href {https://doi.org/https://doi.org/10.1016/j.physc.2005.12.066}
  {\bibfield  {journal} {\bibinfo  {journal} {Physica C: Superconductivity}\
  }\textbf {\bibinfo {volume} {434}},\ \bibinfo {pages} {194} (\bibinfo {year}
  {2006})}\BibitemShut {NoStop}%
\bibitem [{\citenamefont {Janod}\ \emph {et~al.}(1993)\citenamefont {Janod},
  \citenamefont {Junod}, \citenamefont {Graf}, \citenamefont {Wang},
  \citenamefont {Triscone},\ and\ \citenamefont {Muller}}]{janod1993split}%
  \BibitemOpen
  \bibfield  {author} {\bibinfo {author} {\bibfnamefont {E.}~\bibnamefont
  {Janod}}, \bibinfo {author} {\bibfnamefont {A.}~\bibnamefont {Junod}},
  \bibinfo {author} {\bibfnamefont {T.}~\bibnamefont {Graf}}, \bibinfo {author}
  {\bibfnamefont {K.-Q.}\ \bibnamefont {Wang}}, \bibinfo {author}
  {\bibfnamefont {G.}~\bibnamefont {Triscone}},\ and\ \bibinfo {author}
  {\bibfnamefont {J.}~\bibnamefont {Muller}},\ }\bibfield  {title} {\bibinfo
  {title} {Split superconducting transitions in the specific heat and magnetic
  susceptibility of yba2cu3ox versus oxygen content},\ }\href
  {https://doi.org/https://doi.org/10.1016/0921-4534(93)90643-5} {\bibfield
  {journal} {\bibinfo  {journal} {Physica C: Superconductivity}\ }\textbf
  {\bibinfo {volume} {216}},\ \bibinfo {pages} {129} (\bibinfo {year}
  {1993})}\BibitemShut {NoStop}%
\bibitem [{\citenamefont {Khlyustikov}(2011)}]{khlyustikov2011critical}%
  \BibitemOpen
  \bibfield  {author} {\bibinfo {author} {\bibfnamefont {I.~N.}\ \bibnamefont
  {Khlyustikov}},\ }\bibfield  {title} {\bibinfo {title} {{Critical magnetic
  field of surface superconductivity in lead}},\ }\href
  {https://doi.org/10.1134/S1063776111140056} {\bibfield  {journal} {\bibinfo
  {journal} {Journal of Experimental and Theoretical Physics}\ }\textbf
  {\bibinfo {volume} {113}},\ \bibinfo {pages} {1032} (\bibinfo {year}
  {2011})}\BibitemShut {NoStop}%
\bibitem [{\citenamefont {Khlyustikov}(2016)}]{khlyustikov2016surface}%
  \BibitemOpen
  \bibfield  {author} {\bibinfo {author} {\bibfnamefont {I.~N.}\ \bibnamefont
  {Khlyustikov}},\ }\bibfield  {title} {\bibinfo {title} {{Surface
  superconductivity in lead}},\ }\href
  {https://doi.org/10.1134/S1063776116020047} {\bibfield  {journal} {\bibinfo
  {journal} {Journal of Experimental and Theoretical Physics}\ }\textbf
  {\bibinfo {volume} {122}},\ \bibinfo {pages} {328} (\bibinfo {year}
  {2016})}\BibitemShut {NoStop}%
\bibitem [{\citenamefont {Kozhevnikov}\ \emph {et~al.}(2007)\citenamefont
  {Kozhevnikov}, \citenamefont {Bael}, \citenamefont {Sahoo}, \citenamefont
  {Temst}, \citenamefont {Haesendonck}, \citenamefont {Vantomme},\ and\
  \citenamefont {Indekeu}}]{kozhevnikov2007observation}%
  \BibitemOpen
  \bibfield  {author} {\bibinfo {author} {\bibfnamefont {V.~F.}\ \bibnamefont
  {Kozhevnikov}}, \bibinfo {author} {\bibfnamefont {M.~J.~V.}\ \bibnamefont
  {Bael}}, \bibinfo {author} {\bibfnamefont {P.~K.}\ \bibnamefont {Sahoo}},
  \bibinfo {author} {\bibfnamefont {K.}~\bibnamefont {Temst}}, \bibinfo
  {author} {\bibfnamefont {C.~V.}\ \bibnamefont {Haesendonck}}, \bibinfo
  {author} {\bibfnamefont {A.}~\bibnamefont {Vantomme}},\ and\ \bibinfo
  {author} {\bibfnamefont {J.~O.}\ \bibnamefont {Indekeu}},\ }\bibfield
  {title} {\bibinfo {title} {Observation of wetting-like phase transitions in a
  surface-enhanced type-i superconductor},\ }\href
  {https://doi.org/10.1088/1367-2630/9/3/075} {\bibfield  {journal} {\bibinfo
  {journal} {New Journal of Physics}\ }\textbf {\bibinfo {volume} {9}},\
  \bibinfo {pages} {75} (\bibinfo {year} {2007})}\BibitemShut {NoStop}%
\bibitem [{\citenamefont {Khlyustikov}(2021)}]{khlyustikov2021surface}%
  \BibitemOpen
  \bibfield  {author} {\bibinfo {author} {\bibfnamefont {I.~N.}\ \bibnamefont
  {Khlyustikov}},\ }\bibfield  {title} {\bibinfo {title} {{Surface
  Superconductivity of Vanadium}},\ }\href
  {https://doi.org/10.1134/S1063776121030043} {\bibfield  {journal} {\bibinfo
  {journal} {Journal of Experimental and Theoretical Physics}\ }\textbf
  {\bibinfo {volume} {132}},\ \bibinfo {pages} {453} (\bibinfo {year}
  {2021})}\BibitemShut {NoStop}%
\bibitem [{\citenamefont {Mangel}\ \emph {et~al.}(2020)\citenamefont {Mangel},
  \citenamefont {Kapon}, \citenamefont {Blau}, \citenamefont {Golubkov},
  \citenamefont {Gavish},\ and\ \citenamefont
  {Keren}}]{mangel2020stiffnessometer}%
  \BibitemOpen
  \bibfield  {author} {\bibinfo {author} {\bibfnamefont {I.}~\bibnamefont
  {Mangel}}, \bibinfo {author} {\bibfnamefont {I.}~\bibnamefont {Kapon}},
  \bibinfo {author} {\bibfnamefont {N.}~\bibnamefont {Blau}}, \bibinfo {author}
  {\bibfnamefont {K.}~\bibnamefont {Golubkov}}, \bibinfo {author}
  {\bibfnamefont {N.}~\bibnamefont {Gavish}},\ and\ \bibinfo {author}
  {\bibfnamefont {A.}~\bibnamefont {Keren}},\ }\bibfield  {title} {\bibinfo
  {title} {Stiffnessometer: A magnetic-field-free superconducting stiffness
  meter and its application},\ }\href
  {https://doi.org/10.1103/PhysRevB.102.024502} {\bibfield  {journal} {\bibinfo
   {journal} {Phys. Rev. B}\ }\textbf {\bibinfo {volume} {102}},\ \bibinfo
  {pages} {024502} (\bibinfo {year} {2020})}\BibitemShut {NoStop}%
\bibitem [{\citenamefont {Tsindlekht}\ \emph {et~al.}(2004)\citenamefont
  {Tsindlekht}, \citenamefont {Leviev}, \citenamefont {Asulin}, \citenamefont
  {Sharoni}, \citenamefont {Millo}, \citenamefont {Felner}, \citenamefont
  {Paderno}, \citenamefont {Filippov},\ and\ \citenamefont
  {Belogolovskii}}]{tsindlekht2004tunneling}%
  \BibitemOpen
  \bibfield  {author} {\bibinfo {author} {\bibfnamefont {M.~I.}\ \bibnamefont
  {Tsindlekht}}, \bibinfo {author} {\bibfnamefont {G.~I.}\ \bibnamefont
  {Leviev}}, \bibinfo {author} {\bibfnamefont {I.}~\bibnamefont {Asulin}},
  \bibinfo {author} {\bibfnamefont {A.}~\bibnamefont {Sharoni}}, \bibinfo
  {author} {\bibfnamefont {O.}~\bibnamefont {Millo}}, \bibinfo {author}
  {\bibfnamefont {I.}~\bibnamefont {Felner}}, \bibinfo {author} {\bibfnamefont
  {Y.~B.}\ \bibnamefont {Paderno}}, \bibinfo {author} {\bibfnamefont {V.~B.}\
  \bibnamefont {Filippov}},\ and\ \bibinfo {author} {\bibfnamefont {M.~A.}\
  \bibnamefont {Belogolovskii}},\ }\bibfield  {title} {\bibinfo {title}
  {Tunneling and magnetic characteristics of superconducting
  ${\mathrm{zrb}}_{12}$ single crystals},\ }\href
  {https://doi.org/10.1103/PhysRevB.69.212508} {\bibfield  {journal} {\bibinfo
  {journal} {Phys. Rev. B}\ }\textbf {\bibinfo {volume} {69}},\ \bibinfo
  {pages} {212508} (\bibinfo {year} {2004})}\BibitemShut {NoStop}%
\bibitem [{\citenamefont {Belogolovskii}\ \emph {et~al.}(2011)\citenamefont
  {Belogolovskii}, \citenamefont {Felner},\ and\ \citenamefont
  {Shaternik}}]{belogolovskii2010zirconium}%
  \BibitemOpen
  \bibfield  {author} {\bibinfo {author} {\bibfnamefont {M.}~\bibnamefont
  {Belogolovskii}}, \bibinfo {author} {\bibfnamefont {I.}~\bibnamefont
  {Felner}},\ and\ \bibinfo {author} {\bibfnamefont {V.}~\bibnamefont
  {Shaternik}},\ }\bibfield  {title} {\bibinfo {title} {Zirconium dodecaboride,
  a novel superconducting material with enhanced surface characteristics},\
  }in\ \href@noop {} {\emph {\bibinfo {booktitle} {Boron Rich Solids}}},\
  \bibinfo {editor} {edited by\ \bibinfo {editor} {\bibfnamefont
  {N.}~\bibnamefont {Orlovskaya}}\ and\ \bibinfo {editor} {\bibfnamefont
  {M.}~\bibnamefont {Lugovy}}}\ (\bibinfo  {publisher} {Springer Netherlands},\
  \bibinfo {address} {Dordrecht},\ \bibinfo {year} {2011})\ pp.\ \bibinfo
  {pages} {195--206}\BibitemShut {NoStop}%
\bibitem [{\citenamefont {Khasanov}\ \emph {et~al.}(2005)\citenamefont
  {Khasanov}, \citenamefont {Di~Castro}, \citenamefont {Belogolovskii},
  \citenamefont {Paderno}, \citenamefont {Filippov}, \citenamefont
  {Br\"utsch},\ and\ \citenamefont {Keller}}]{khasanov2005anomalous}%
  \BibitemOpen
  \bibfield  {author} {\bibinfo {author} {\bibfnamefont {R.}~\bibnamefont
  {Khasanov}}, \bibinfo {author} {\bibfnamefont {D.}~\bibnamefont {Di~Castro}},
  \bibinfo {author} {\bibfnamefont {M.}~\bibnamefont {Belogolovskii}}, \bibinfo
  {author} {\bibfnamefont {Y.}~\bibnamefont {Paderno}}, \bibinfo {author}
  {\bibfnamefont {V.}~\bibnamefont {Filippov}}, \bibinfo {author}
  {\bibfnamefont {R.}~\bibnamefont {Br\"utsch}},\ and\ \bibinfo {author}
  {\bibfnamefont {H.}~\bibnamefont {Keller}},\ }\bibfield  {title} {\bibinfo
  {title} {Anomalous electron-phonon coupling probed on the surface of
  superconductor $\mathrm{Zr}{\mathrm{b}}_{12}$},\ }\href
  {https://doi.org/10.1103/PhysRevB.72.224509} {\bibfield  {journal} {\bibinfo
  {journal} {Phys. Rev. B}\ }\textbf {\bibinfo {volume} {72}},\ \bibinfo
  {pages} {224509} (\bibinfo {year} {2005})}\BibitemShut {NoStop}%
\bibitem [{\citenamefont {Nakada}\ \emph {et~al.}(1996)\citenamefont {Nakada},
  \citenamefont {Fujita}, \citenamefont {Dresselhaus},\ and\ \citenamefont
  {Dresselhaus}}]{nakada1996edge}%
  \BibitemOpen
  \bibfield  {author} {\bibinfo {author} {\bibfnamefont {K.}~\bibnamefont
  {Nakada}}, \bibinfo {author} {\bibfnamefont {M.}~\bibnamefont {Fujita}},
  \bibinfo {author} {\bibfnamefont {G.}~\bibnamefont {Dresselhaus}},\ and\
  \bibinfo {author} {\bibfnamefont {M.~S.}\ \bibnamefont {Dresselhaus}},\
  }\bibfield  {title} {\bibinfo {title} {Edge state in graphene ribbons:
  Nanometer size effect and edge shape dependence},\ }\href@noop {} {\bibfield
  {journal} {\bibinfo  {journal} {Physical Review B}\ }\textbf {\bibinfo
  {volume} {54}},\ \bibinfo {pages} {17954} (\bibinfo {year}
  {1996})}\BibitemShut {NoStop}%
\bibitem [{\citenamefont {Fujita}\ \emph {et~al.}(1996)\citenamefont {Fujita},
  \citenamefont {Wakabayashi}, \citenamefont {Nakada},\ and\ \citenamefont
  {Kusakabe}}]{fujita1996peculiar}%
  \BibitemOpen
  \bibfield  {author} {\bibinfo {author} {\bibfnamefont {M.}~\bibnamefont
  {Fujita}}, \bibinfo {author} {\bibfnamefont {K.}~\bibnamefont {Wakabayashi}},
  \bibinfo {author} {\bibfnamefont {K.}~\bibnamefont {Nakada}},\ and\ \bibinfo
  {author} {\bibfnamefont {K.}~\bibnamefont {Kusakabe}},\ }\bibfield  {title}
  {\bibinfo {title} {Peculiar localized state at zigzag graphite edge},\
  }\href@noop {} {\bibfield  {journal} {\bibinfo  {journal} {Journal of the
  Physical Society of Japan}\ }\textbf {\bibinfo {volume} {65}},\ \bibinfo
  {pages} {1920} (\bibinfo {year} {1996})}\BibitemShut {NoStop}%
\bibitem [{\citenamefont {Wakabayashi}\ \emph {et~al.}(1999)\citenamefont
  {Wakabayashi}, \citenamefont {Fujita}, \citenamefont {Ajiki},\ and\
  \citenamefont {Sigrist}}]{wakabayashi1999electronic}%
  \BibitemOpen
  \bibfield  {author} {\bibinfo {author} {\bibfnamefont {K.}~\bibnamefont
  {Wakabayashi}}, \bibinfo {author} {\bibfnamefont {M.}~\bibnamefont {Fujita}},
  \bibinfo {author} {\bibfnamefont {H.}~\bibnamefont {Ajiki}},\ and\ \bibinfo
  {author} {\bibfnamefont {M.}~\bibnamefont {Sigrist}},\ }\bibfield  {title}
  {\bibinfo {title} {Electronic and magnetic properties of nanographite
  ribbons},\ }\href@noop {} {\bibfield  {journal} {\bibinfo  {journal}
  {Physical Review B}\ }\textbf {\bibinfo {volume} {59}},\ \bibinfo {pages}
  {8271} (\bibinfo {year} {1999})}\BibitemShut {NoStop}%
\bibitem [{\citenamefont {Wakabayashi}\ \emph {et~al.}(2010)\citenamefont
  {Wakabayashi}, \citenamefont {Sasaki}, \citenamefont {Nakanishi},\ and\
  \citenamefont {Enoki}}]{Wakabayashi_2010}%
  \BibitemOpen
  \bibfield  {author} {\bibinfo {author} {\bibfnamefont {K.}~\bibnamefont
  {Wakabayashi}}, \bibinfo {author} {\bibfnamefont {K.-i.}\ \bibnamefont
  {Sasaki}}, \bibinfo {author} {\bibfnamefont {T.}~\bibnamefont {Nakanishi}},\
  and\ \bibinfo {author} {\bibfnamefont {T.}~\bibnamefont {Enoki}},\ }\bibfield
   {title} {\bibinfo {title} {Electronic states of graphene nanoribbons and
  analytical solutions},\ }\href@noop {} {\bibfield  {journal} {\bibinfo
  {journal} {Science and technology of advanced materials}\ }\textbf {\bibinfo
  {volume} {11}},\ \bibinfo {pages} {054504} (\bibinfo {year}
  {2010})}\BibitemShut {NoStop}%
\bibitem [{\citenamefont {Kobayashi}\ \emph {et~al.}(2005)\citenamefont
  {Kobayashi}, \citenamefont {Fukui}, \citenamefont {Enoki}, \citenamefont
  {Kusakabe},\ and\ \citenamefont {Kaburagi}}]{PhysRevB.71.193406}%
  \BibitemOpen
  \bibfield  {author} {\bibinfo {author} {\bibfnamefont {Y.}~\bibnamefont
  {Kobayashi}}, \bibinfo {author} {\bibfnamefont {K.-i.}\ \bibnamefont
  {Fukui}}, \bibinfo {author} {\bibfnamefont {T.}~\bibnamefont {Enoki}},
  \bibinfo {author} {\bibfnamefont {K.}~\bibnamefont {Kusakabe}},\ and\
  \bibinfo {author} {\bibfnamefont {Y.}~\bibnamefont {Kaburagi}},\ }\bibfield
  {title} {\bibinfo {title} {Observation of zigzag and armchair edges of
  graphite using scanning tunneling microscopy and spectroscopy},\ }\href
  {https://doi.org/10.1103/PhysRevB.71.193406} {\bibfield  {journal} {\bibinfo
  {journal} {Phys. Rev. B}\ }\textbf {\bibinfo {volume} {71}},\ \bibinfo
  {pages} {193406} (\bibinfo {year} {2005})}\BibitemShut {NoStop}%
\bibitem [{\citenamefont {Sugawara}\ \emph {et~al.}(2006)\citenamefont
  {Sugawara}, \citenamefont {Sato}, \citenamefont {Souma}, \citenamefont
  {Takahashi},\ and\ \citenamefont {Suematsu}}]{PhysRevB.73.045124}%
  \BibitemOpen
  \bibfield  {author} {\bibinfo {author} {\bibfnamefont {K.}~\bibnamefont
  {Sugawara}}, \bibinfo {author} {\bibfnamefont {T.}~\bibnamefont {Sato}},
  \bibinfo {author} {\bibfnamefont {S.}~\bibnamefont {Souma}}, \bibinfo
  {author} {\bibfnamefont {T.}~\bibnamefont {Takahashi}},\ and\ \bibinfo
  {author} {\bibfnamefont {H.}~\bibnamefont {Suematsu}},\ }\bibfield  {title}
  {\bibinfo {title} {Fermi surface and edge-localized states in graphite
  studied by high-resolution angle-resolved photoemission spectroscopy},\
  }\href {https://doi.org/10.1103/PhysRevB.73.045124} {\bibfield  {journal}
  {\bibinfo  {journal} {Phys. Rev. B}\ }\textbf {\bibinfo {volume} {73}},\
  \bibinfo {pages} {045124} (\bibinfo {year} {2006})}\BibitemShut {NoStop}%
\bibitem [{\citenamefont {Shtanko}\ and\ \citenamefont
  {Levitov}(2018)}]{shtanko2018robustness}%
  \BibitemOpen
  \bibfield  {author} {\bibinfo {author} {\bibfnamefont {O.}~\bibnamefont
  {Shtanko}}\ and\ \bibinfo {author} {\bibfnamefont {L.}~\bibnamefont
  {Levitov}},\ }\bibfield  {title} {\bibinfo {title} {Robustness and
  universality of surface states in dirac materials},\ }\href@noop {}
  {\bibfield  {journal} {\bibinfo  {journal} {Proceedings of the National
  Academy of Sciences}\ }\textbf {\bibinfo {volume} {115}},\ \bibinfo {pages}
  {5908} (\bibinfo {year} {2018})}\BibitemShut {NoStop}%
\bibitem [{\citenamefont {Pangburn}\ \emph {et~al.}(2022)\citenamefont
  {Pangburn}, \citenamefont {Haurie}, \citenamefont {Cr{\'e}pieux},
  \citenamefont {Awoga}, \citenamefont {Black-Schaffer}, \citenamefont
  {P{\'e}pin},\ and\ \citenamefont {Bena}}]{pangburn2022superconductivity}%
  \BibitemOpen
  \bibfield  {author} {\bibinfo {author} {\bibfnamefont {E.}~\bibnamefont
  {Pangburn}}, \bibinfo {author} {\bibfnamefont {L.}~\bibnamefont {Haurie}},
  \bibinfo {author} {\bibfnamefont {A.}~\bibnamefont {Cr{\'e}pieux}}, \bibinfo
  {author} {\bibfnamefont {O.~A.}\ \bibnamefont {Awoga}}, \bibinfo {author}
  {\bibfnamefont {A.~M.}\ \bibnamefont {Black-Schaffer}}, \bibinfo {author}
  {\bibfnamefont {C.}~\bibnamefont {P{\'e}pin}},\ and\ \bibinfo {author}
  {\bibfnamefont {C.}~\bibnamefont {Bena}},\ }\bibfield  {title} {\bibinfo
  {title} {Superconductivity in monolayer and few-layer graphene: I. review of
  possible pairing symmetries and basic electronic properties},\ }\href@noop {}
  {\bibfield  {journal} {\bibinfo  {journal} {arXiv preprint arXiv:2211.05146}\
  } (\bibinfo {year} {2022})}\BibitemShut {NoStop}%
\bibitem [{\citenamefont {Barkman}\ \emph {et~al.}(2019)\citenamefont
  {Barkman}, \citenamefont {Samoilenka},\ and\ \citenamefont
  {Babaev}}]{barkman2019surface}%
  \BibitemOpen
  \bibfield  {author} {\bibinfo {author} {\bibfnamefont {M.}~\bibnamefont
  {Barkman}}, \bibinfo {author} {\bibfnamefont {A.}~\bibnamefont
  {Samoilenka}},\ and\ \bibinfo {author} {\bibfnamefont {E.}~\bibnamefont
  {Babaev}},\ }\bibfield  {title} {\bibinfo {title} {Surface pair-density-wave
  superconducting and superfluid states},\ }\href@noop {} {\bibfield  {journal}
  {\bibinfo  {journal} {Physical review letters}\ }\textbf {\bibinfo {volume}
  {122}},\ \bibinfo {pages} {165302} (\bibinfo {year} {2019})}\BibitemShut
  {NoStop}%
\bibitem [{\citenamefont {Saroka}\ \emph {et~al.}(2017)\citenamefont {Saroka},
  \citenamefont {Shuba},\ and\ \citenamefont {Portnoi}}]{Saroka2017Optics}%
  \BibitemOpen
  \bibfield  {author} {\bibinfo {author} {\bibfnamefont {V.}~\bibnamefont
  {Saroka}}, \bibinfo {author} {\bibfnamefont {M.}~\bibnamefont {Shuba}},\ and\
  \bibinfo {author} {\bibfnamefont {M.}~\bibnamefont {Portnoi}},\ }\bibfield
  {title} {\bibinfo {title} {Optical selection rules of zigzag graphene
  nanoribbons},\ }\href@noop {} {\bibfield  {journal} {\bibinfo  {journal}
  {Physical Review B}\ }\textbf {\bibinfo {volume} {95}},\ \bibinfo {pages}
  {155438} (\bibinfo {year} {2017})}\BibitemShut {NoStop}%
\bibitem [{\citenamefont {Talkachov}\ and\ \citenamefont
  {Babaev}(2022)}]{talkachov2022wave}%
  \BibitemOpen
  \bibfield  {author} {\bibinfo {author} {\bibfnamefont {A.}~\bibnamefont
  {Talkachov}}\ and\ \bibinfo {author} {\bibfnamefont {E.}~\bibnamefont
  {Babaev}},\ }\bibfield  {title} {\bibinfo {title} {Wave functions and edge
  states in rectangular honeycomb lattices revisited: nanoflakes, armchair and
  zigzag nanoribbons and nanotubes},\ }\href@noop {} {\bibfield  {journal}
  {\bibinfo  {journal} {arXiv preprint arXiv:2208.08555 Physical Review B in
  print}\ } (\bibinfo {year} {2022})}\BibitemShut {NoStop}%
\bibitem [{\citenamefont {Wakabayashi}\ and\ \citenamefont
  {Dutta}(2012)}]{wakabayashi2012nanoscale}%
  \BibitemOpen
  \bibfield  {author} {\bibinfo {author} {\bibfnamefont {K.}~\bibnamefont
  {Wakabayashi}}\ and\ \bibinfo {author} {\bibfnamefont {S.}~\bibnamefont
  {Dutta}},\ }\bibfield  {title} {\bibinfo {title} {Nanoscale and edge effect
  on electronic properties of graphene},\ }\href@noop {} {\bibfield  {journal}
  {\bibinfo  {journal} {Solid state communications}\ }\textbf {\bibinfo
  {volume} {152}},\ \bibinfo {pages} {1420} (\bibinfo {year}
  {2012})}\BibitemShut {NoStop}%
\bibitem [{\citenamefont {Onipko}\ and\ \citenamefont
  {Malysheva}(2018)}]{Onipko2018Revisit}%
  \BibitemOpen
  \bibfield  {author} {\bibinfo {author} {\bibfnamefont {A.}~\bibnamefont
  {Onipko}}\ and\ \bibinfo {author} {\bibfnamefont {L.}~\bibnamefont
  {Malysheva}},\ }\bibfield  {title} {\bibinfo {title} {Electron spectrum of
  graphene macromolecule revisited},\ }\href@noop {} {\bibfield  {journal}
  {\bibinfo  {journal} {Physica Status Solidi (B)}\ }\textbf {\bibinfo {volume}
  {255}},\ \bibinfo {pages} {1700248} (\bibinfo {year} {2018})}\BibitemShut
  {NoStop}%
\bibitem [{\citenamefont {Zheng}\ \emph {et~al.}(2007)\citenamefont {Zheng},
  \citenamefont {Wang}, \citenamefont {Luo}, \citenamefont {Shi},\ and\
  \citenamefont {Chen}}]{zheng2007analytical}%
  \BibitemOpen
  \bibfield  {author} {\bibinfo {author} {\bibfnamefont {H.}~\bibnamefont
  {Zheng}}, \bibinfo {author} {\bibfnamefont {Z.}~\bibnamefont {Wang}},
  \bibinfo {author} {\bibfnamefont {T.}~\bibnamefont {Luo}}, \bibinfo {author}
  {\bibfnamefont {Q.}~\bibnamefont {Shi}},\ and\ \bibinfo {author}
  {\bibfnamefont {J.}~\bibnamefont {Chen}},\ }\bibfield  {title} {\bibinfo
  {title} {Analytical study of electronic structure in armchair graphene
  nanoribbons},\ }\href@noop {} {\bibfield  {journal} {\bibinfo  {journal}
  {Phys. Rev. B}\ }\textbf {\bibinfo {volume} {75}},\ \bibinfo {pages} {165414}
  (\bibinfo {year} {2007})}\BibitemShut {NoStop}%
\bibitem [{\citenamefont {Wei{\ss}e}\ \emph {et~al.}(2006)\citenamefont
  {Wei{\ss}e}, \citenamefont {Wellein}, \citenamefont {Alvermann},\ and\
  \citenamefont {Fehske}}]{weisse2006kernel}%
  \BibitemOpen
  \bibfield  {author} {\bibinfo {author} {\bibfnamefont {A.}~\bibnamefont
  {Wei{\ss}e}}, \bibinfo {author} {\bibfnamefont {G.}~\bibnamefont {Wellein}},
  \bibinfo {author} {\bibfnamefont {A.}~\bibnamefont {Alvermann}},\ and\
  \bibinfo {author} {\bibfnamefont {H.}~\bibnamefont {Fehske}},\ }\bibfield
  {title} {\bibinfo {title} {The kernel polynomial method},\ }\href@noop {}
  {\bibfield  {journal} {\bibinfo  {journal} {Reviews of modern physics}\
  }\textbf {\bibinfo {volume} {78}},\ \bibinfo {pages} {275} (\bibinfo {year}
  {2006})}\BibitemShut {NoStop}%
\bibitem [{\citenamefont {Covaci}\ \emph {et~al.}(2010)\citenamefont {Covaci},
  \citenamefont {Peeters},\ and\ \citenamefont {Berciu}}]{covaci2010efficient}%
  \BibitemOpen
  \bibfield  {author} {\bibinfo {author} {\bibfnamefont {L.}~\bibnamefont
  {Covaci}}, \bibinfo {author} {\bibfnamefont {F.}~\bibnamefont {Peeters}},\
  and\ \bibinfo {author} {\bibfnamefont {M.}~\bibnamefont {Berciu}},\
  }\bibfield  {title} {\bibinfo {title} {Efficient numerical approach to
  inhomogeneous superconductivity: the chebyshev-bogoliubov--de gennes
  method},\ }\href@noop {} {\bibfield  {journal} {\bibinfo  {journal} {Physical
  review letters}\ }\textbf {\bibinfo {volume} {105}},\ \bibinfo {pages}
  {167006} (\bibinfo {year} {2010})}\BibitemShut {NoStop}%
\bibitem [{\citenamefont {Nagai}\ \emph {et~al.}(2012)\citenamefont {Nagai},
  \citenamefont {Ota},\ and\ \citenamefont {Machida}}]{nagai2012efficient}%
  \BibitemOpen
  \bibfield  {author} {\bibinfo {author} {\bibfnamefont {Y.}~\bibnamefont
  {Nagai}}, \bibinfo {author} {\bibfnamefont {Y.}~\bibnamefont {Ota}},\ and\
  \bibinfo {author} {\bibfnamefont {M.}~\bibnamefont {Machida}},\ }\bibfield
  {title} {\bibinfo {title} {Efficient numerical self-consistent mean-field
  approach for fermionic many-body systems by polynomial expansion on spectral
  density},\ }\href@noop {} {\bibfield  {journal} {\bibinfo  {journal} {Journal
  of the Physical Society of Japan}\ }\textbf {\bibinfo {volume} {81}},\
  \bibinfo {pages} {024710} (\bibinfo {year} {2012})}\BibitemShut {NoStop}%
\bibitem [{\citenamefont {Gibbs}(1899)}]{gibbs1899fourier}%
  \BibitemOpen
  \bibfield  {author} {\bibinfo {author} {\bibfnamefont {J.~W.}\ \bibnamefont
  {Gibbs}},\ }\bibfield  {title} {\bibinfo {title} {Fourier's series},\
  }\href@noop {} {\bibfield  {journal} {\bibinfo  {journal} {Nature}\ }\textbf
  {\bibinfo {volume} {59}},\ \bibinfo {pages} {606} (\bibinfo {year}
  {1899})}\BibitemShut {NoStop}%
\bibitem [{\citenamefont {Wilbraham}(1848)}]{wilbraham1848cambridge}%
  \BibitemOpen
  \bibfield  {author} {\bibinfo {author} {\bibfnamefont {H.}~\bibnamefont
  {Wilbraham}},\ }\bibfield  {title} {\bibinfo {title} {Cambridge and dublin
  math},\ }\href@noop {} {\bibfield  {journal} {\bibinfo  {journal} {J}\
  }\textbf {\bibinfo {volume} {3}},\ \bibinfo {pages} {198} (\bibinfo {year}
  {1848})}\BibitemShut {NoStop}%
\bibitem [{\citenamefont {Zhu}(2016)}]{zhu2016bogoliubov}%
  \BibitemOpen
  \bibfield  {author} {\bibinfo {author} {\bibfnamefont {J.-X.}\ \bibnamefont
  {Zhu}},\ }\href@noop {} {\emph {\bibinfo {title} {Bogoliubov-de Gennes method
  and its applications}}},\ Vol.\ \bibinfo {volume} {924}\ (\bibinfo
  {publisher} {Springer},\ \bibinfo {year} {2016})\BibitemShut {NoStop}%
\end{thebibliography}%

\end{document}